\documentclass{amsart}
\pdfoutput=1

\usepackage{natbib}
\usepackage{amsthm}
\newcommand{\half}{\frac{1}{2}}
\usepackage[pdftex]{graphicx,color}

\usepackage{myMac-sig-alt}

\newcommand{\sqfr}{\mbox{\tt SqFree}}
\newcommand{\Dif}[1]{\partial_{#1}}
\newcommand{\grad}{\nabla}

\newcommand{\EVB}{\mbox{\tt EV}}

\newcommand{\PV}{Plantinga \& Vegter}

\newcommand{\fP}{{\mathfrak{P}}}      
\newcommand{\V}{\mbox{\sc Zero}}       
\newcommand{\lc}{\mathop{lc}}       
\newcommand{\otm}{\ot{m}}       %
\newcommand{\otn}{\ot{n}}       %
\newcommand{\olzeta}{\ol{\zeta}}  
\DeclareMathOperator{\Height}{Height}
\DeclareMathOperator{\interior}{int}
\newcommand{\ot}[1]{\widetilde{#1}}

\bibpunct{[}{]}{;}{a}{,}{,}

\begin{document}
\DeclareGraphicsExtensions{.pdftex}
\DeclareGraphicsRule{.pdftex}{pdf}{*}{}


\title[Isotopic Meshing of Singular Algebraic Curves]{Complete Subdivision Algorithms, II:\\
    Isotopic Meshing of Singular Algebraic Curves}

\author{Michael Burr}
\address{Courant Institute, NYU 251 Mercer Street, NY, NY 10012.}
\curraddr{Mathematics Department, Fordham University, Bronx, NY 10458.}
\email{burr@cims.nyu.edu}
\thanks{Michael Burr's work is partially supported by NSF Grant CCF-0728977.}

\author{Sung Woo Choi}
\address{Department of Mathematics, Duksung Women's University, Seoul 132-714, Korea.}
\email{swchoi@ducksung.ac.kr}

\author{Ben Galehouse}
\address{Courant Institute, NYU 251 Mercer Street, NY, NY 10012.}
\curraddr{Max-Planck-Institut f\"ur Informtik Department 1: Algorithms and Complexity, 66123 Saarbr\"ucken, Germany.}
\thanks{Ben Galehouse's work is partially supported by the DoE under contract DE-FG02-00ER25053.}
\email{bgalehou@mpi-inf.mpg.de}

\author{Chee Yap}
\address{Korea Institute of Advanced Study, Seoul, Korea and Courant Institute, NYU 251 Mercer Street, NY, NY 10012.}
\thanks{Chee Yap's work is supported by NSF Grant CCF-0728977.}
\email{yap@cs.nyu.edu}

\begin{abstract}
Given a real valued function $f(X,Y)$, a box region $B_0\ib\RR^2$ and $\vareps>0$,
we want to compute an $\vareps$-isotopic polygonal
approximation to the restriction of the curve $S=f^{-1}(0)$ $=\set{p\in\RR^2: f(p)=0}$ to $B_0$.
We focus on subdivision algorithms because of their adaptive complexity
and ease of implementation.
\PV\ gave a numerical subdivision algorithm that
is exact when the curve $S$ is bounded and non-singular.
They used a computational model that relied
only on function evaluation and interval arithmetic.
We generalize their algorithm to any bounded (but possibly
non-simply connected) region that does not contain singularities of $S$.
With this generalization as a subroutine,
we provide a method to detect isolated algebraic singularities and
their branching degree.  This appears to be
the first complete {\em purely numerical} method to compute
isotopic approximations of algebraic curves with isolated singularities.

\end{abstract}

\maketitle

\section{Introduction}

Given $\vareps>0$, a box region $B_0\ib\RR^2$
and a real valued function $f:\RR^2\to\RR$, we want to
compute a polygonal approximation $P$ to the restriction of the
implicit curve $S=f^{-1}(0)$ to $B_0$ (where $f^{-1}(0)=\set{p\in\RR^2: f(p)=0}$).
The approximation $P$ must be
(1) ``topologically correct'' and (2) ``$\vareps$-close'' to $S\cap B_0$.
We use the standard interpretation of requirement (2),
that $d(P,S\cap B_0)\le \vareps$
where $d(\cdot,\cdot)$ is the Hausdorff distance on compact sets.
In recent years, it has become accepted
\citep{boissonnat+4:meshing:06}
to interpret requirement (1) to mean that
$P$ is isotopic to $S\cap B_0$, which we denote by $P\approx S\cap B_0$.
This means that we not only require that $P$ and $S\cap B_0$ are
homeomorphic, but also require that they are embedded in $\RR^2$
``in the same way''.
This means that the two embeddings can be continuously deformed to each other.
E.g., if $S\cap B_0$ consists of two disjoint ovals, these can be embedded
in $\RR^2$ as two ovals exterior to each other,
or as two nested ovals.  Isotopy, but not homeomorphism,
requires $P$ to respect this distinction.
There is a stronger notion of isotopy called \dt{ambient isotopy}
(see the definition in Section 4).
We use this stronger notion in this paper
(but, for simplicity, we still say ``isotopy'').
See \citep[p.~183]{boissonnat+4:meshing:06} for a
discussion of the connection between ambient and plain isotopy.
In this paper, we focus mainly on topological correctness since
achieving $\vareps$-closeness is not an issue
for our particular subdivision approach
(cf.~\citep[pp.~213-4]{boissonnat+4:meshing:06}).
This amounts to setting $\vareps=\infty$.

We may call the preceding problem the \dt{2-D implicit meshing problem}.
The term ``meshing'' comes from the corresponding problem
in 3-D:  given $\vareps>0$ and an implicit surface $S:f(X,Y,Z)=0$,
we want to construct a triangular mesh $M$ such that $d(M,S)\le \vareps$
and $M\approx S$.  It is interesting (see \citep{burr-sharma-yap:eval:09})
to identify the 1-D meshing with the well-known problem
of real root isolation and refinement for a real function $f(X)$.

The \dt{algebraic approach} and the \dt{numerical approach}
constitute two extremes of a spectrum
among the approaches to most computational problems on curves and surfaces.
Algebraic methods can clearly solve most problems in this area,
e.g., by an application of the general theory of cylindrical algebraic
decomposition (CAD) \citep{basu-pollack-roy:bk}.
Purely algebraic methods, however, are generally not considered practical,
even in the plane
(e.g., \citep{hong:plane-curves:96,seidel-wolpert:topology:05}),
but efficient solutions have been achieved
for special cases such as intersecting quadrics
in 3-D \citep{schoemer-wolpert:quad:06}.
At the other end of the spectrum,
the numerical approaches emphasize approximation and iteration.
An important class of such algorithms is the class of
\dt{subdivision algorithms} which can be viewed as
a generalization of binary search.
Such algorithms are practical in two
senses: they are easy to implement and their
complexity is more adaptive with respect to the input instance \citep{yap:subdiv1:06}.
Another key feature of subdivision algorithms
is that they are ``localized'', meaning that
we can restrict our computation to
some region of interest.

Besides the algebraic and numerical approaches, there is another
approach that might be called the \dt{geometric approach} in which
we postulate an abstract computational model with certain
(geometric) primitives (e.g., shoot an ray or decide if a point is in a cell).
When implementing these geometric algorithms, one
must still choose an algebraic or numeric implementation of these
primitives.  Implementations can also use a hybrid of
algebraic and numeric techniques.

Unfortunately, numerical methods
seldom have global correctness guarantees.  The most famous example
is the Marching Cube algorithm \citep{marching-cube}.
Many authors have tried to improve the correctness of subdivision algorithms
(e.g., \citep{stander-hart:topology:97}).
So far, such efforts have succeeded under one of the following situations:
    \bitem
    \item (A0) Requiring niceness assumptions such
    as being non-singular or Morse.
    \item (A1) Invoking algebraic techniques such as
        resultant computations or manipulations of algebraic numbers.
    \eitem

It is clear that (A0) should be avoided.
Generally, we call a method ``complete'' if the method
is correct without any (A0) type restrictions.
But many incomplete algorithms (e.g., Marching cube)
are quite useful in practice.
We want to avoid (A1) conditions because algebraic
manipulations are harder to implement and such techniques
are relatively expensive and non-adaptive \citep{yap:subdiv1:06}.
The complete removal of (A0) type restrictions
is the major open problem
faced by purely numerical approaches to meshing.
Thus, \citep[p.~187]{boissonnat+4:meshing:06} state
that ``{\em meshing in the vicinity of singularities is a
difficult open problem and an active area of research}''.
Most of the techniques described
in their survey are unable to handle singularities.
It should be evident that this
open problem has an implicit requirement to avoid the use
of (A1) techniques.

For example, the subdivision meshing algorithm of
\citep{plantinga:thesis:06,plantinga-vegter:isotopic:04}
requires the non-singularity and the boundedness of curves and surfaces, (A0) assumptions.
The subdivision algorithm
of \citep{seidel-wolpert:topology:05}
requires\footnote{
    Their paper is subtitled ``Exploiting a little more Geometry and a little
    less Algebra'' which speaks to our concerns with (A1).
} the computation of resultants, an (A1) technique.
We thus classify \citep{seidel-wolpert:topology:05} as a hybrid approach
that combines numerical and algebraic techniques.
Prior to our work, we are not aware of any meshing algorithm that
can handle singularities without resorting to resultant computations.
In general, hybrid methods offer
considerable promise (e.g., \citep{hong:plane-curves:96}).
This is part of a growing trend to employ numerical techniques
to speed up algebraic computations.

Some of our recent work addresses the above (A0)/(A1) concerns:
in \citep{yap:subdiv1:06}, we gave a complete numerical approach for
determining tangential Bezier curve intersections;
in \citep{cheng-gao-yap:triangular:07}, we numerically solve
zero-dimensional triangular systems without any
``regularity'' requirements on the systems;
in \citep{burr-sharma-yap:eval:09}, we provide
numerical root isolation in the presence of multiple zeros;
and \citep{burr-krahmer-yap:continuousAmort:09} provides
one of the first non-probabilistic adaptive analyses
of an evaluation-based real root isolation algorithm.
These last two papers address the 1-D analogue of the \PV\ Algorithm.
The philosophy behind all of these papers is the design and
analysis of complete numerical methods based on approximations,
iteration and adaptive methods.  Topological
exactness is achieved using suitable algebraic bounds,
ranging from classical root separation bounds
to evaluation bounds and geometric separation bounds.
We stress that the worst-case complexity of adaptive
algorithms ought not to be the chief criterion for
evaluating the usefulness of these algorithms:
for the majority of inputs, these algorithms terminate fast.
Zero bounds are only used as stopping
criteria for iteration in the algorithms,
and simple estimates for them can be computed easily.
Computing such bounds does not mean that we compute resultants,
even though their justification depends on resultant theory.
The present paper continues this line of investigation.

The recent collection
\citep[Chapter 5]{boissonnat+4:meshing:06}
reviews the current algorithmic literature in meshing in 2- and 3-D:
the subdivision approach is represented by
the \PV\
algorithm as well as by Snyder's earlier
approach based on parametrizability
\citep{snyder:interval:92,snyder:generative:bk}.
The subdivision algorithm of \PV\ is remarkable in the
following sense: even though it is globally isotopic,
it does not guarantee isotopy of the curve
within each cell of the subdivision.   In contrast,
Snyder's subdivision approach
\citep{snyder:interval:92,snyder:generative:bk}
requires the correct isotopy type in each cell.
Indeed, because of this, the algorithm
is incomplete \citep[p.~195]{boissonnat+4:meshing:06}.

Among geometric approaches to meshing, we have the
point sampling approach as represented by
\citep{boissonnat-oudot:lipschitz:06,cheng+3:sampling:04},
the Morse Theory approach as represented
by \citep{stander-hart:topology:97,boissonnat+2:meshing-topology:04}
and the sweepline approach \citep{mourrain-tecourt:surface:05}.
Note that the sweepline approach naturally corresponds to
the algebraic operation of projection; therefore, its implementation
is often purely algebraic.  The idea of the sampling approach
is to reduce meshing of a surface $S$ to computing
the Delaunay triangulation of a sufficiently dense
set of sample points on $S$
\citep[p.~201--213]{boissonnat+4:meshing:06}.
To obtain such sample points,
\citep{cheng+3:sampling:04} need a primitive operation that
amounts to solving a system of equations involving $f$ and its derivatives.
\citep{boissonnat-oudot:lipschitz:06} need
a primitive for intersecting the surface with a Voronoi edge.
These sample points are algebraic, and implementing the primitives
exactly would require strong algebraic techniques.
But exact implementation does not seem to be justified
for these applications, and so we are faced
with an implementation gap that shows well-known
non-robustness issues.  For restrictions and open problems
in sampling approaches,
see \citep[p.~227--229]{boissonnat+4:meshing:06}.
In contrast, the computational primitives
needed by subdivision approachs work directly with
bigfloats, with modest requirements on $f$.

This paper presents a purely numerical subdivision
method for meshing algebraic curves with isolated singularities.
In a certain sense, this is the most general geometric
situation because, by \refPro{finitesubset}, reduced algebraic
curves have only isolated singularities.
Our starting point is the algorithm of
\PV\ \citep{plantinga-vegter:isotopic:04,plantinga:thesis:06}
for implicit meshing of curves.  It is important to understand
the computational model of \PV\ which is also used in this paper.
Two capabilities are assumed with regards to $f(X,Y)$:
\bitem
\item
(i) Sign evaluation of $f(p)$ at dyadic points $p$.
\item
(ii) $f$ is $C^1$ and we can evaluate the interval
analogues (i.e., box functions) of $f, \pdiff{f}{X}, \pdiff{f}{Y}$
on dyadic intervals.
\eitem
Note that the Marching Cube algorithm only requires capability (i).
Let the \dt{class $PV$} denote the set of
all real functions $f:\RR^2\to\RR$ for which capabilities
(i) and (ii) are available.
Many common functions of analysis
(e.g., the elementary functions \citep{du-yap:hyper:06})
belong to $PV$; thus, the approach of \PV\ admits a more
general setting than algebraic curves.

\subsection{Overview of Paper}
\bitem
\item Section 2 establishes some basic terminology and
    recalls facts about the singularities of algebraic sets.
\item In Sections 3 and 4,
    we extend the \PV\ algorithm to compute
    an isotopic approximation of the curve $S=f^{-1}(0)$ restricted
    to a ``nice region'' that need not be simply connected.
    $S$ may have singularities outside $R$ and we only need $f\in PV$.
\item In Section 5, we provide the algebraic evaluation bounds
    necessary for meshing singular curves.
\item In Section 6,
    we provide a subdivision method to isolate all the
    singularities of a square-free integer polynomial $f(X,Y)$.
\item In Section 7,
    given a box $B$ containing an isolated singularity $p$,
    we provide a method to compute the branching degree of $p$.
\item In Section 8,
    we finally present the overall algorithm
    to compute the isotopic polygonal approximation.
\item We conclude in Section 9.
\eitem

\section{Basic Terminology and Algebraic Facts}

Let $\FF\as \ZZ\left[\half\right]=\set{m2^n: m,n\in \ZZ}$
be the set of \dt{dyadic numbers}.
All of our numerical computations are performed in a straightforward manner using $\FF$.
There are many well-known implementations of arithmetic
on such numbers, and, in this case, they known as \dt{bigfloats}.
In short, our computational model is not based on some
abstract capability whose implementation may reveal gaps
that lead to well-known non-robustness issues.

For $S\ib\RR$, let $\intbox S$ be the set of closed intervals $[a,b]$
with endpoints in $S$, $a,b\in S$.
We write $\intbox S^n$ for $(\intbox S)^n$.
In particular, $\intbox\FF$ is the set of dyadic intervals,
and $\intbox\RR^n$ is the set of $n$-boxes.
The \dt{width} of an interval $I=[a,b]$ is $w(I)\as b-a$.
The width and diameter, respectively, of an $n$-box $B=\prod_{i=1}^n [a_i,b_i]$ is
$w(B)\as \min\set{b_i-a_i: i=1\dd n}$ and
$d(B)\as \max\set{b_i-a_i: i=1\dd n}$.
The boundary of a set $S\ib\RR$ is denoted $\partial S$.
If $f:\RR^n\to\RR$ and $S\ib\RR$, then $f(S)\as\set{f(x): x\in S}$.
A function $\intbox f:\intbox\FF^n\to\intbox\FF$
is a \dt{box function} for $f$ provided (i) $f(B)\ib \intbox f(B)$
and (ii) if $B_0\ip B_1\ip\cdots$ with $\lim_i B_i= p$ then
$\lim_i \intbox f(B_i)=f(p)$.  We regard the limit of intervals
in terms of the limits of their endpoints.
We say $f\in PV$ if $f\in C^1$ (has
continuous first derivatives), there
is an algorithm to determine $\sign(f(p))$
for $p\in\FF^n$ and $\intbox f$ and the corresponding functions
for the derivatives of $f$ are computable in $\intbox\FF$.
In this paper,
we only consider box functions for the two dimensional case.

We only consider boxes of the form $B=I\times J$
where $I, J$ are dyadic intervals.
As in \PV, all our boxes will be squares, $w(I)=w(J)=w(B)$.
Our algorithms work in the slightly more general setting
where all boxes have aspect ratios at most $2$
(see \citep{lin-yap:cxy:09} for a proof).
The boundary $\partial B$ of $B$ is divided into four \dt{sides}
and four \dt{corners}.
Note that the `sides/corners' terminology for boxes should not be confused
with the `edges/vertices' terminology which we reserve for the straightline
graph $G=(V,E)$ which is the approximation to our curve.
We \dt{split} a box $B$ by subdividing it into 4 subboxes of equal widths.
These subboxes are the \dt{children} of $B$ and each has width $\frac{1}{2}w(B)$.
Starting with $B_0$,
the child-parent relationships obtained by an arbitrary sequence of
splits yields a \dt{quadtree} rooted at $B_0$.
Two distinct boxes $B, B'$ of a quadtree are \dt{neighbors} if their boundaries overlap:
$B\cap B'$ is a line segment, but not a single point.
Segments of the form $B\cap B'$ are called \dt{interior segments}.
If $B$ is a box whose side $s$ is part of the boundary of $B_0$, then
we call $s$ a \dt{boundary segment} and $B$ a \dt{boundary box}.
Moreover, each side of a box is divided into one or more segments.

Some of the regions of interest in this paper will not be squares and may not even be simply connected.
To ensure that our algorithm continues to work in this case,
we insist that the region comes from a subdivision.
I.e., we insist that there exists a square $B_0$, a subdivision of $B_0$ and
a collection of boxes in the subdivision such that the region of interest is the union of this
collection of boxes.
Although it is not necessary for our algorithms, in most implementations, it is
easiest to maintain a subdivision of $B_0$ where each box is labeled as either region or
complement.  The union of the boxes labeled ``region'' will be written $R_0$ (or $R_0(T)$ when we stress the subdivision tree)
and is the region of interest.
We reserve the notation $R_0$ for non-square regions and use $B_0$ for square regions.

{\bf Algebraic Facts.}
Let $\DD$ be a unique factorization domain (UFD)
and $f,g\in \DD[\bfX]=\DD[X_1\dd X_n]$
where $\bfX=(X_1\dd X_n)$.  We say $f, g$ are \dt{similar}
if there exist $a,b\in \DD\setminus\{0\}$ such that $af=bg$, and we
write this relationship as $f\sim g$.  Otherwise, $f$ and $g$ are \dt{dissimilar}.
The \dt{square-free part} of $f$ is defined as
    \beql{sqfr}
    \sqfr(f) \as \frac{f}{\gcd(f, \Dif{X_1}f \dd \Dif{X_n}f)}
    \eeql
where $\Dif{X_i}$ indicates differentiation
with respect to $X_i$.  $f$ is said to be \dt{square-free} if $\sqfr(f)=f$.
From \refeq{sqfr} we see that
computing $\sqfr(f)$ from $f$ involves only rational operations of $\DD$.
As the gradient of $f$ is $\grad f= (\Dif{X_1}f \dd \Dif{X_n}f)$, we may
also write $\gcd(f,\grad f)$ for $\gcd(f, \Dif{X_1}f \dd \Dif{X_n}f)$.
See \citep[Chap.~2]{yap:algebra:bk} for standard conventions concerning
$\gcd$.

Let $k$ be an algebraically closed field.
For $S\ib k[\bfX]=k[X_1\dd X_n]$,
let $\V(S)\as \set{p\in k^n: f(p)=0 \mbox{\rm\ for all } f\in S}$
denote the \dt{zero set} of $S$.  A zero set is also known
as a \dt{variety}.
The \dt{singular points} of $\V(f)$ are defined to be the points
where $\grad\sqfr(f)=0$.

In 1-dimension, it is well-known that
a square-free polynomial $f\in\ZZ[X]$
has no singularities (i.e., multiple zeros).  We now recall
two generalizations of this result that will be necessary in the
remainder of the paper.
See \citep{hartshorne:bk,clo1:bk,harris:alge-geom:bk} for similar results.

\bproT{\protect{\citep[Ex.14.3]{harris:alge-geom:bk}}}{finitesubset}
The singular points of any variety form a proper subvariety.
\eproT

This proposition is critical in our paper, because it
implies that if $f\in \RR[X,Y]$ is square-free,
then the singular points are a proper subvariety of a union of
curves and hence must be a finite set of points.
Thus, we only need to handle isolated singularities.

\bproT{Algebraic Sard Lemma \protect{\citep[Prop.14.4]{harris:alge-geom:bk}}}{sard}
Let $f:X\rightarrow Y$ be any surjective regular map of varieties defined over a
field $k$ of characteristic $0$.
Then there exists a nonempty open subset $U\ib Y$
such that for any smooth point
$p\in f^{-1}(U)\cap X_{sm}$ in the inverse image of $U$, the
differential $df_p$ is surjective.
\eproT

Here, $X_{sm}$ denotes the set of smooth points of variety $X$.
The open sets refer to the Zariski topology.
The condition that the differential $df_p$ is surjective is equivalent to
insisting that the Jacobian of $f$ has the same rank as the dimension of $Y$.
The situation that we consider
is $f:\RR^2\rightarrow\RR$ where $f$ is a polynomial.
In this case, since the image is one dimensional,
the condition that $df_p$ is surjective
reduces to the condition that $\grad f(p)\not=0$.
Every point in $\RR^2=X$ is smooth and $\RR\setminus U$ is
only a finite set.
Hence, there are only a finite number of level sets, parametrized by $h$,
where $\V(f(X,Y)-h)$ has a singular point.

\section{Algorithm of \PV}
\label{basepv}

\begin{figure}[htb]
    \begin{center}
	\scalebox{0.5}{
    	  \input{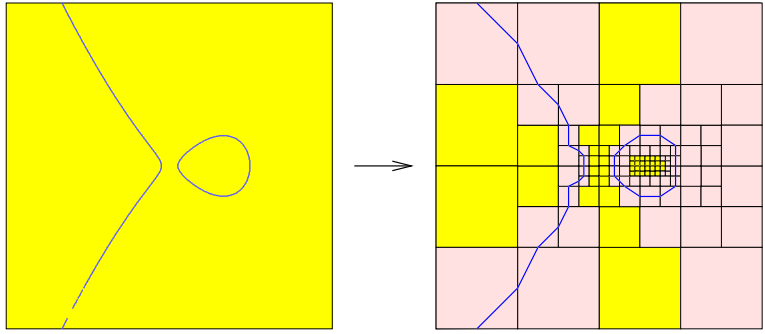}}
    \caption{(a) Original curve $Y^2-X^2+X^3+0.2=0$, (b) Isotopic approximation
    }
    \label{fig:bigpic}
    \end{center}
    \end{figure}

First, we recall the \PV\ algorithm:
Given $\vareps>0$, a bounded and nonsingular curve $f^{-1}(0)$
for $f:\RR^2\to\RR$ and a bounding box $B_0\in\intbox\FF^2$,
they compute a polygonal curve $P$.
The correctness statement is as follows:
{\em if $f^{-1}(0)\ib B_0$, then $P$ is an
$\vareps$-approximation to the curve $S=f^{-1}(0)$, i.e.,
$d(P,S)\le\vareps$ and $P\approx S$.}
For simplicity, they focus on topological correctness:
$P\approx S$, since it is easy to refine the subdivision
to achieve $d(P,S)\le \vareps$.
The curves in \refFig{bigpic} are output from our
implementation of the \PV\ algorithm as reported in \citep{lin-yap:cxy:09}.
The value of $\vareps$ is small in \refFig{bigpic}(a), while
$\vareps=\infty$ in \refFig{bigpic}(b).

The algorithm is based on two simple predicates on boxes $B$:
\begin{itemize}
\item  Predicate $C_0(B)$ holds if $0\not\in \intbox f(B)$.
\item  Predicate $C_1(B)$ holds if
$0\not\in \left(\intbox \pdiff{f}{X}(B)\right)^2 + \left(\intbox \pdiff{f}{Y}(B)\right)^2$.
\end{itemize}
These predicates are easily implemented for $f\in PV$.  If $C_0(B)$ holds, then
the curve $S$ does not intersect $B$.  If $C_1(B)$ holds,
then the gradient of $f$ is never zero in $B$ and the gradient
vectors approximately point in the same direction.  Note that if
$B$ satisfies $C_1$, then, by recalling the parent-child relationship,
any child of $B$ also satisfies $C_1$.

The input box $B_0$ is a dyadic square
and the output will be an undirected graph $G=(V,E)$ where
each vertex $v\in V$ is a dyadic point, $v\in \FF^2$.
In fact, $G$ represents a straightline planar graph
that is a polygonal approximation of $S$.

The algorithm has 3 phases, where Phase $i$ ($i=1,2,3$) is associated with
a queue $Q_i$ containing boxes.
Initially, $Q_1=\set{B_0}$, and $Q_2=Q_3=\es$.
When $Q_i$ is empty, proceed to Phase $i+1$.
\begin{itemize}

\item PHASE 1: SUBDIVISION.
While $Q_1$ is non-empty, remove some $B$ from $Q_1$,
and perform the following:
If $C_0(B)$ holds, discard $B$.
If $C_1(B)$ holds, insert $B$ into $Q_2$.
Otherwise, split $B$ into four subboxes and
insert them into $Q_1$.

\item PHASE 2: BALANCING.
This phase ``balances'' the subdivision, where
a subdivision is \dt{balanced} if the widths of any two neighboring
boxes differ by at most a factor of $2$.
Queue $Q_2$ is a min-priority queue, where the width of
a box serves as its priority.
While $Q_2$ is non-empty, remove the min-element $B$ from $Q_2$,
and perform the following:
For each $B$-neighbor $B'$ in $Q_2$ with width more than twice the width
of $B$, remove $B'$ from $Q_2$ and split $B'$.
Insert each child $B''$ of $B'$ into $Q_2$ provided $C_0(B'')$ does not hold.
$B''$ might be a new neighbor of $B$ and $B''$
might be split subsequently.
Finally, when every neighbor of $B$ is at most twice the width of $B$, insert $B$ into $Q_3$.

\item PHASE 3: CONSTRUCTION.
This phase constructs the graph $G=(V,E)$.
Initially, the boxes in $Q_3$ are unmarked.
While $Q_3$ is non-empty, remove any $B$ from $Q_3$
and mark it. Now construct the set $V(B)$ of its vertices.
For each $B$-neighbor $B'$, if $B'$ is marked, retrieve any vertex $v$
on the side $B\cap B'$, and put $v$ into $V(B)$.
If $B'$ is unmarked, evaluate
the sign of $f(p)f(q)$ where $p,q$ are endpoints of the segment $B\cap B'$.
If $f(p)f(q)<0$, create a vertex $v= (p+q)/2$ for the graph $G$ and put $v$ into $V(B)$.
Note that if $f(p)=0$ for any corner $p$, treat $f(p)$
as positive; in effect, this is an infinitesimal perturbation at $p$.
It can be shown that $|V(B)|\in\set{0, 2, 4}$.
If $|V(B)|=2$, put the edge $(p,q)$ into $G$ to connect the vertices
in $V(B)$.
If $|V(B)|=4$, it can be shown that one side of $B$
contains two vertices $a,b$.  Introduce two edges into $E$
to connect each of $a,b$ to the remaining two vertices.
The requirement that these two edges are non-crossing ensures
that the connection is unique
(see \citep{plantinga-vegter:isotopic:04,plantinga:thesis:06}).
\end{itemize}

The output graph $G=(V,E)$ can be viewed as a straightline graph,
decomposed into a collection $P=P(G)$
of closed polygons.
In the following, we simply use $G$ in place of $P$ as the polygonal approximation.

\section{Extension of \PV\ }

\subsection{The Na\"{i}ve Extension of \PV}

\begin{figure}[htb]
    \begin{center}
	\scalebox{0.35}{
    	  \input{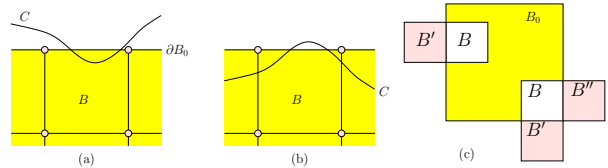}}
    \caption{
    (a) incursion,
    (b) excursion,
    (c) boundary boxes and their complements
    }
    \label{fig:boundaryBox}
    \end{center}
    \end{figure}

The correctness statement of \PV\ requires $B_0$
to be a bounding box for the curve $f^{-1}(0)$.
The power of subdivision methods
comes from their ability to adaptively analyze local data.
Choosing the initial box to be a bounding box gives
up this power of local analysis; also, by insisting that the curve is bounded,
the types of possible curves are severely restricted.
As a first attempt,
one could na\"{i}vely attempt to run the \PV\ Algorithm
starting with an arbitrary box
and then ask the question ``In what sense is the output $G$ correct?''
Intuitively, $G$ should be isotopic to $f^{-1}(0)\cap B_0$,
but \PV\ did not discuss this issue.
The algorithm certainly cannot handle the case when the curve $S=f^{-1}(0)$ has tangential
but non-crossing intersections \citep{yap:subdiv1:06} with $\partial B_0$.
If we assume that there are only transversal intersections,
we still face two problems:
if the curve $S$ (locally) enters and exits $\partial B_0$ by
visiting only one box $B\ib B_0$,
the above algorithm would fail to detect this small component.
Such an error is called an undetected \dt{incursion},
as illustrated in \refFig{boundaryBox}(a).
Conversely, the curve $S$ might escape undetected from $B_0$.
Such an error is called an undetected \dt{excursion},
as illustrated in \refFig{boundaryBox}(b).

These errors cause the final approximation to be incorrect,
since $S\cap B_0$ may not have the same number of components as $G$.
Incursions account for components of $S\cap B_0$ that are undetected by the \PV\ algorithm,
and excursions account for connected components of $G$ that are not connected in $S\cap B_0$.
A single box may contain several incursions and excursions.
The excursion case is more troubling because there is
no guarantee that $C_1$ will hold
in the complement of $B_0$; this means
that the connected components of $G$ which are not connected
in $S\cap B_0$ might, in fact, not even be connected in $S$.
If we choose $B_0$ large enough, such errors cannot arise,
but this approach gives up the power of adaptivity and localization
which subdivision methods possess.
If $S$ has singularities, making $B_0$ large may not be an option.
In this paper, we avoid any ``largeness'' assumption on $B_0$.

The preceding issues arise because the
\PV\ algorithm focuses only on the parity of the endpoints of an edge of the subdivision.
They prove that multiple intersections can be removed
by applying a suitable isotopy to
move small features into a neighboring box.
We reproduce their results here in order to examine them.
In the following, we fix a balanced quadtree $T$ rooted at a box $B_0$.
Let $V(T)$ denote the collection of boxes at the leaves of $T$.
These boxes constitute a partition of $B_0$.
We also assume that $C_0(B)$ or $C_1(B)$ holds at each $B\in V(T)$.
The subdivision constructed through the \PV\ algorithm has this property.
The results in this section do not need the tree $T$ to be balanced
because this constraint is only used in the construction of the isotopic curve.

In PHASE 3 of the \PV\ algorithm, we noted that
if $f^{-1}(0)$ passes though a corner of the subdivision,
then they treat that point as positive.
This can be done via an infinitesimal perturbation of $f$:
we call this $\wtf$ the \dt{standard perturbation} of $f$
with respect to $T$.
More precisely, $\wtf$ is defined to be a function that agrees with $f$
everywhere except in infinitesimal neighborhoods of
corners of boxes where $f$ is $0$; in such neighborhoods, $\wtf$ is positive
at the corner of the box.
Clearly, the definition of $\wtf$ depends on $T$.
In general, $\wt{S}=\wtf^{-1}(0)$ is isotopic to $S=f^{-1}(0)$,
but in the case where $B_0$ is not a bounding box for the curve,
it can happen that
$\wt{S}\cap B_0$ is not isotopic to $S\cap B_0$.
This difference is discussed more throughly in Section \ref{WCS}.
All of our correctness statements are about $\wt{S}$ and not $S$.
However, for the sake of simplicity in this section, we continue to use the notations $S$
and $f$ instead of $\wt{S}$ and $\wtf$.

\begin{figure}[htb]
    \begin{center}
	\scalebox{0.28}{
    	  \input{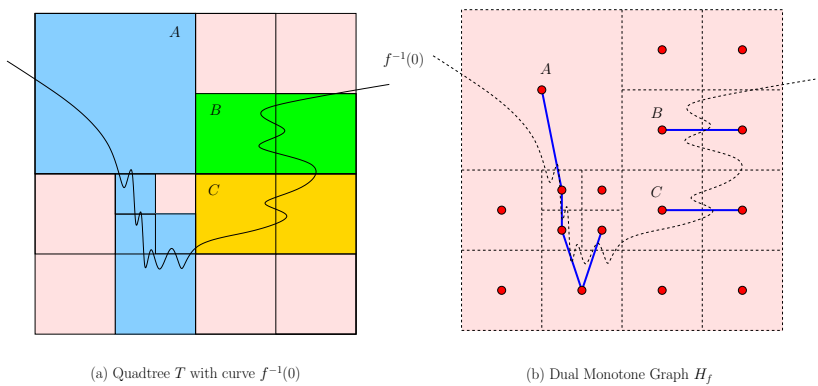}}
    \caption{A quadtree $T$ and its dual graph induced by $f^{-1}(0)$.
    }
    \label{fig:globalIsotopyDual2}
    \end{center}
    \end{figure}

Consider the \dt{dual graph} $H=H(T)$ whose vertex set is $V(T)$
and edges connect pairs of neighboring boxes.
Thus, $B,B'\in V(T)$ are connected by an edge in $H$
iff $s=B\cap B'$ is a segment, but not a point.
We are interested in the subgraph $H_f=H_f(T)$ of $H$
in which $B,B'$ are connected by an edge iff
the segment $s=B\cap B'$ intersects the
curve $f^{-1}(0)$ more than once.  We further focus on the
non-trivial connected components of $H_f$, i.e., those components
consisting of more than one box.
In \refFig{globalIsotopyDual2}, the graph $H_f$ has three
non-trivial connected components ($A, B, C$).
The union of all of the boxes corresponding
to a non-trivial connected component of $H_f$ is called
a \dt{monotone region}, and $H_f$ itself will
be known as the \dt{monotone graph}.
In \refFig{globalIsotopyDual2}(a), we have filled each monotone
region ($A, B, C$) with a distinct tint.

\begin{figure}[htb]
    \begin{center}
	\scalebox{0.25}{
    	  \input{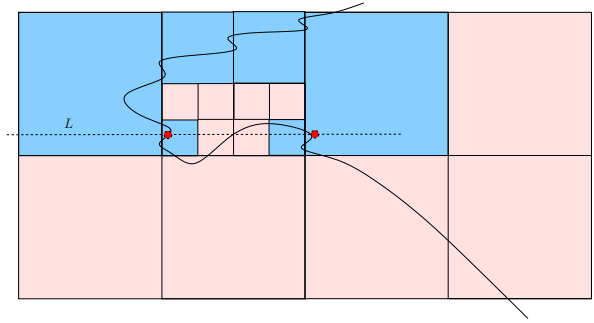}}
    \caption{An $X$-monotone region that is not globally parametrizable in the $Y$-direction.
    }
    \label{fig:localparam}
    \end{center}
    \end{figure}

Let us explain the monotone terminology.
Each monotone region $M$ can be uniquely classified as
a \dt{$Y$-monotone} or \dt{$X$-monotone} region in the following sense:
we say $M$ is \dt{$Y$-monotone relative to $f$}
if for all $B,B'\ib M$ where $(B,B')$ is an edge of $H_f$,
then $B\cap B'$ is a segment orthogonal to the $Y$-direction
(i.e., $s$ is horizontal).  Moreover,
    \beql{0notin}
    0\notin \pdiff{f}{Y}(B\cup B').
    \eeql
We call such a rectangular region $B\cup B'$ a \dt{local neighborhood} of the monotone region.
Thus \refeq{0notin} implies that the curve $f^{-1}(0)$ restricted to each local neighborhood
of a $Y$-monotone region is
parametrizable in the $X$-direction (i.e., each vertical line intersects
the curve at most once).  This is a ``local property'' because the curve
$f^{-1}(0)$ might not be parametrizable in the $X$-direction in the entire $Y$-monotone region.
See \refFig{localparam} for an example of a $X$-monotone region in which
the curve $f^{-1}(0)$ is not globally parametrizable
in the $Y$-direction: a horizontal line $L$ intersets the curve twice in the monotone region.
The curve is parametrizable, however, for connected components of $L$ with respect to the $X$-monotone region.
The definition of a $X$-monotone regions is
similar, after exchanging the roles of $X$ \& $Y$ and the roles of horizontal \& vertical.

\begin{figure}[htb]
    \begin{center}
	\scalebox{0.30}{
    	  \input{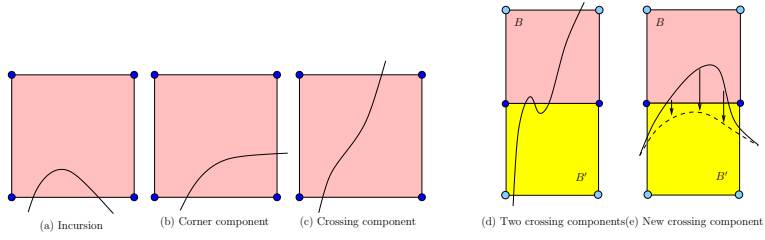}}
    \caption{Types of components
	and their isotopic transformation in a local neighborhood.
    }
    \label{fig:components}
    \end{center}
    \end{figure}

Let $\mathcal{C}$ be a connected component of $f^{-1}(0)\cap B$.
We classify $\mathcal{C}$ into three types:
if both endpoints of $\mathcal{C}$ lie on one side of $B$,
then $\mathcal{C}$ is an incursion as discussed above.
If the endpoints of $\mathcal{C}$ lie on two adjacent
sides of $B$, it is called a \dt{corner component}.
Finally, if the endpoints of $\mathcal{C}$
lie on two opposite sides of $B$, it is called a \dt{crossing component}.
These three kinds of components are illustrated in \refFig{components}(a-c).
Note that there can be no other types of components in $B$.
In particular, $B$ cannot contain an isolated component
since $C_1(B)$ holds.
The next lemma also restricts the number of crossing components.

Let us define the notion of isotopy used in this paper.
An \dt{isotopy} is a continuous function
    $I:\RR^2\times[0,1]\to\RR^2$
such that the following two conditions hold:
(1) $I(\cdot,0)$ is the identity
and
(2) $I(\cdot,t)$ is a homeomorphism for each $t\in [0,1]$.
If, in addition, there is a set $B\ib\RR^2$ such that for all $t\in [0,1]$,
$I(\cdot,t)$ acts as the identity on the complement of $B$, i.e.,
$I(p,t)=p$ for all $p\not\in B$, then we call $I$
an \dt{isotopy on $B$}, and write the isotopy as $I:B\times [0,1]\to B$.
Two sets $S,S'\ib\RR$ are said to be \dt{isotopic}, denoted
$S\approx S'$,
if there exists an isotopy $I$ such that $I(S,1)=S'$.
As promised in the introduction,
our notion of isotopy is the ``ambient'' sort
\citep[p.~183]{boissonnat+4:meshing:06}.
Observe that if $I$ is an isotopy on $B$,
then for any continuous function $f:B\to\RR$,
$f^{-1}(0)$ and $(f\circ I^{-1}(\cdot,t))^{-1}(0)$
are isotopic.  We state without proof the following
lemma which is somewhat implicit
\citep{plantinga-vegter:isotopic:04,plantinga:thesis:06}:

\bleml{PVlem}
Let $(B, B')$ be an edge of the monotone graph $H_f$.
Assume their shared segment $s=B\cap B'$ is horizontal.
\begin{enumerate}
\item
$B\cup B'$ is $Y$-monotone relative to $f$
\item
There is at most one crossing component in each of $B$ and $B'$.
If there is one crossing component in each of $B$ and $B'$ then
these two components are part of one single crossing component
of $B\cup B'$.
\item
Let $f^{-1}(0)$ intersect $s$ at two consecutive points $p_1$ and $p_2$.
Then $p_1$ and $p_2$ are connected by a curve component
$X(p_1,p_2)\ib f^{-1}(0)$ that lies  entirely within $B$ or
entirely within $B'$.
\item
There is an isotopy $I$ on $B\cup B'$
that reduces the number of intersections (counted
with multiplicities) on $s$ so that
$(f\circ I(\cdot,1)^{-1})^{-1}(0)$ intersects $s$ in $2$ fewer times
than $f^{-1}(0)$, and $B\cup B'$ remains
$Y$-monotone relative to $f\circ I(\cdot,t)^{-1}$ for all $t$.
\end{enumerate}
\eleml

Part (2) of this Lemma follows from the fact that
$C_1(B\cup B')$ holds.
The situation where $B$ and $B'$ both has a crossing
component in illustrated in \refFig{components}(d).
The isotopy $I$ on $B\cup B'$ constructed in part (4)
cannot decrease the number of crossing components,
but may increase the number by one,
as illustrated in \refFig{components}(e).
The general idea is to repeatedly apply the transformation given
by such an isotopy $I$ until
the curve $f^{-1}(0)$ composed with the inverses of the
isotopies intersects the segment $s$
at most once (see \citep{lin-yap:cxy:09} for a generalization
under a weaker predicate than $C_1$).
The isotopies can easily be chosen to preserve monotonicity
by keeping the appropriate coordinate fixed throughout the isotopy,
i.e., the $X$-coordinate can be kept fixed
for an isotopy on a $Y$-monotone regions.
This prepares us for an induction because after each isotopic
transformation, monotonicity is preserved.
To develop the appropriate global correctness statement,
we formulate the sense in which these isotopies interact:

\bleml{pv:correctness}
Let $H$ be the dual graph to the subdivision given by the tree $T$,
i.e., there is an edge connecting boxes $B_1$ and $B_2$
iff they are neighbors.
Let $H_f$ be the subgraph of $H$ where $B_1$ and $B_2$ are
connected by an edge of $H$
and $f^{-1}(0)$ intersects their shared segment more than once.
Let $D$ be a connected component of $H_f$, and
$R$ be the union of the boxes appearing in $D$.
There exists an isotopy $I$ on $R$ such that
if $f_1= f\circ I(\cdot,1)^{-1}$ then
$f^{-1}(0)\cap R\approx f_1^{-1}(0)\cap R$ and
$f_1^{-1}(0)$ intersects all segments between boxes of $D$ at most once.
\eleml
Note, in particular, that boxes which correspond to
isolated components of $H_f$ are boxes where $f^{-1}(0)$ intersects
each of its sides at most once.
The upshot of this argument may be summarized by the following theorem, implicit in \PV:

\bthml{normalized}
Let $T$ be the subdivision tree at the end of the \PV\ algorithm.
Then there exists an isotopy $I$ on $B_0$
such that $\wtf^{-1}(0)\cap B_0\approx f_1^{-1}(0)\cap B_0$,
where $f_1:=\wtf\circ I(\cdot,1)^{-1}$.  Moreover, $f_1^{-1}(0)$ intersects each
interior segment of $T$ at most once.
\ethml

A function $f_1$ as given by the conditions of this
theorem, is said to be \dt{normalized} relative to $T$.
It is important to realize that the normalized curve $f_1^{-1}(0)$
may intersect the boundary segments of $T$ more than once.
The reason for this is that we can only apply the transformation of
\refLem{PVlem} to interior segments;
we cannot apply the transformation to boundary segments without changing
the topology of the restriction of the curve to $B_0$.
Therefore, since \PV\ focus entirely on the parity across segments,
the graph $G$ constructed by the \PV\ algorithm may not be
isotopic to $f_1^{-1}(0)\cap B_0$.
We note that although the isotopies constructed here might not preserve the $C_1$ condition
since they are only guaranteed to preserve monotonicity,
it will be useful to recall that these boxes were derived from ones where $C_1$ did hold.

One solution for ensuring that $G$ satisfies $G\approx S\cap B_0$
is to ensure that $S$ intersects each boundary segment at most once.
A sufficient condition is that either $C_0$ or $C_1$ holds on the edge,
using the corresponding properties for the one-dimensional version of \PV's algorithm
(cf.~\citep{burr-sharma-yap:eval:09}).
Achieving this sufficiency condition requires that we determine the topology of $S$ on the boundary of $B_0$,
including tangental intersections.
This solution is also inefficient because it may
require frequent subdivisions near the
boundary to resolve fine features; in
higher dimensions, this solution must also be recursively applied to lower dimensional boxes.
These issues also arise in Snyder's approach \citep{snyder:interval:92}.

\subsection{The Enlarged Region Solution}

We now provide an alternative solution which is closer to the
spirit of the \PV\ approach of exploiting isotopy.
This solution avoids determining the exact boundary topology as well as
making any largness asssumptions on $B_0$.
We wish to slightly enlarge $B_0$ so that
incursions and excursions can be removed by an appropriate isotopy.
The basic idea is that, in addition to subdividing $B_0$, we find a
slightly larger region $B_0'$ which is the union of $B_0$
with a ``collar'' of squares around $B_0$.  We construct a straightline
approximation $G^+$ that will be isotopic to the curve restricted to some
expansion $B_0^+$ of $B_0$ into this collar.  This is weaker than
saying $G^+$ is isotopic to $f^{-1}(0)\cap B_0$, but it has two
major advantages: it allows our algorithm to terminate faster and it does not
require $f^{-1}(0)$ to intersect $\partial B_0$ in any special way.

Call a box $B\ib B_0$ a \dt{boundary box} if $\partial B$
intersects $\partial B_0$.
Let $B$ be such a box.  If $B$ does not share a corner with $B_0$,
then it has a unique \dt{complementary box} $B'$ such
that $B'$ has the same width as $B$,
the interiors of $B'$ and $B_0$ are disjoint, and
$\partial B' \cap \partial B_0 = \partial B \cap \partial B_0$.
$B$ and $B'$ are called \dt{partners} of each other.
If $B$ shares a corner with $B_0$, then
it determines two complementary boxes $B',B''$.
See \refFig{boundaryBox}(c).
We insist that the collar is formed from complementary boxes
all of which satisfy either $C_0$ or $C_1$.
It is easy to adapt \PV's algorithm so that this situation occurs.
With this property,
excursions from $B_0$ or incursions into $B_0$ can be limited to
to the collar region.

Let $B$ be
a boundary box with complementary box $B'$.
If the complementary box satisfies $C_0$, then the curve does not intersect it.
Therefore, there can be no incursions into or excursions from $B$.
If the complementary box satisfies $C_1$ and there is an excursion,
then the isotopy presented in \PV\ shows that the approximation
constructed by the na\"{i}ve \PV\ algorithm is correct in $B\cup B'^+$,
for some region $B\ib B'^+\ib B'\cup B$.
The case that is left to consider is when there is an incursion.
This case is harder because the isotopy constructed by \PV\ would
remove the component of the incursion, but that would result in an error.
In the spirit of \PV's algorithm, we consider the sign pattern of the corners of complementary boxes.

    \begin{figure}[htb]
    \begin{center}
	\scalebox{0.3}{
    	  \input{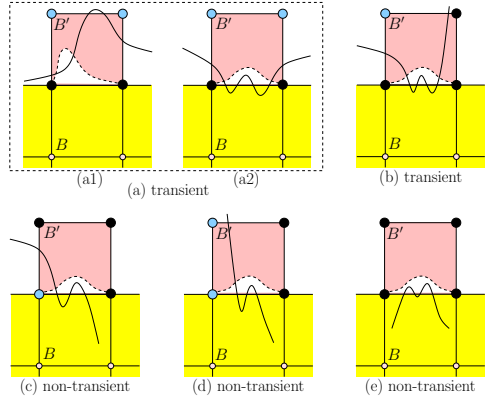}}
    \caption{Classification of complementary boxes according
    to sign of $f$ at the corners.}
    \label{fig:transient}
    \end{center}
    \end{figure}

Up to symmetry and sign flips, $B'$ can be put into
one of five types based solely on the sign of
$f$ at the corners of $B'$, (a)-(e).
This is illustrated in \refFig{transient}.
Note that the ``alternate sign pattern'' does not appear because
this pattern cannot be
$C_1$ as shown by \PV.

We draw two instances of Types (a) to indicate the two possible
``dispositions'' of the curve $\wtf^{-1}(0)$ in $B'$: In (a1)
the curve makes no incursion into $B$,
but (a2) represents at least one incursion into $B$.
These two dispositions give rise to distinct
isotopy types for the curve $\wtf^{-1}(0)\cap B$.
Similarly, type (b) has two possible dispositions
(but we only indicate the case where there
is an incursion).  Types (c), (d), and (e) do not
have analogous dispositions.
Because of this difference in dispositions, we further classify Types (a) and (b) as
\dt{transient}; the other types are called \dt{non-transient}.

{\bf Some Intuition.}
We show how the complementary boxes are used to yield
a correctness statement.  Suppose $B'$ is a complementary box
whose partner is $B$.  Let the straightline $G=(V,E)$,
when restricted to $B$, be denoted $G\cap B$.  Note that
$G\cap B$ has at most two edges.
We would like to claim that $\wtf^{-1}(0)\cap B$ is isotopic to $G\cap B$.
This is evidently false for Types (a) and (b) because
of the two dispositions discussed above; but even for Type (c),
this claim can be false because $\wtf^{-1}(0)\cap B$ may have an arbitrary
number of components due to incursions/excursions,
as illustrated in \refFig{transient}(c).

    \begin{figure}[htb]
    \begin{center}
	\scalebox{0.3}{
    	  \input{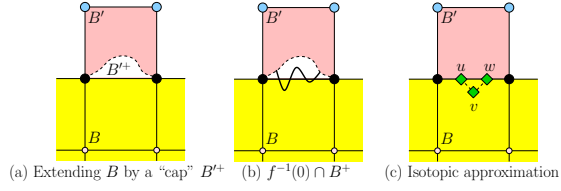}}
    \caption{Representing the transient type: extension of a boundary box by a cap
    }
    \label{fig:isotopy}
    \end{center}
    \end{figure}

To remedy the situation for Types (a) and (b),
we first expand $B$ into a slightly larger region $B^+\ib B\cup B'$.
The actual expansion $B'^+\as B^+\setminus B$
is represented in Figures \ref{fig:transient}, \ref{fig:isotopy}(a,b) by the white region
demarcated by a dashed curve and we call $B'^+$ a \dt{cap}.
Up to isotopy, the incursions into $B^+$
in Types (a1) and (a2) are equivalent to a single incursion
of the curve $\wtf^{-1}(0)$ into $B^+$ from $B'$, as
shown in \refFig{isotopy}(b).
In the graph $G=(V,E)$, we represent this incursion by
adding three vertices $u,v,w$ to $V$,
with $u$ and $w$ on the side $B\cap B'$,
and with $v$ between $u$ and $w$, but slightly to the interior of $B$.
We also add the edges $(u,v), (v,w)$ to $E$.
This is illustrated in \refFig{isotopy}(c).
Let $G^+$ denote the \dt{augmented straightline graph}
with these additional vertices and edges for
transient complementary boxes.

Types (a) and $(b)$  are important because they include the
situation where $f^{-1}(0)$ makes
a non-crossing tangential intersection with the boundary of $B_0$.
Detecting this situation is expensive, and provides a main
motivation for exploiting isotopy.
There is a similar expansion $B^+$ of $B$ into $B'$ for type (c),
as indicated by dashed curves in \refFig{transient}(c).
In this case, we do not need to augment the graph $G=(V,E)$ with any new
vertices or edges, because $G$ already contains an edge representing
this component in $B$.

We capture the above intuitive explanations as a lemma:

\bleml{local}
Assume that $\wtf$ is normalized relative to a quadtree $T$.
Let $B\ib B_0$ be a boundary box and let $B'$ be its partner
(a complementary box) that satisfies $C_1$ or $C_0$.
Then there exists a region $B^+$ such that:
\\(i) $B^+$ is an expansion of $B$ into $B'$:
        $$B\ib B^+ \ib B\cup B'$$
    and the boundary of $B^+$ consists of two connected
    parts $\partial_1(B^+)$ and $\partial_2(B^+)$:
    $\partial_1(B^+)$ is the union of three non-boundary sides of $B$
    and $\partial_2(B^+)$ is either the fourth side of $B$ or a curve in the interior of $B'$
    (in \refFig{transient}, it is a dashed curve).
    In particular $B^+\cap\partial(B')\ib B\cap B'$.
\\(ii) The augmented straightline graph $G^+=(V^+,E^+)$ when restricted to $B$
    is isotopic to $\wtf^{-1}(0)\cap B^+$, i.e.,
        $$G^+\cap B \approx \wtf^{-1}(0)\cap B^+.$$
    Moreover, the isotopy $I$ on $B^+$ that witnesses
    this graph isotopy can be chosen to
    be the identity on $\partial_1(B^+)$.
\eleml
\bpf
Note first that even though $\wtf$ has been normalized,
it came from the situation where all boxes of $T$ satisfy $C_0$ or $C_1$.
This, in turn, implies that the properties of \refLem{PVlem} apply to
incursions and excursions and that the isotopies chosen in the normalization
process can be chosen to be the identity on these incursions and excursions.
Note that $G^+\cap B$ has between $0$ and $3$ components.
This is because at most two edges can appear in $G\cap B$ according
to the \PV\ construction rules, and in Types (a) and (b), the augmented
graph $G^+$ has an additional component.
We now consider each type in turn, and, in each case, define
the expansion $B^+$ of $B$.  These expansions
are illustrated in \refFig{transient}(a)-(e).

For Type (e), only an excursion is possible.
If there is an excursion, we can expand
$B$ into $B^+$ to ensure that $\wtf^{-1}(0)$ never intersects $\partial_2(B^+)$.

For Types (c) and (d), the $\wtf^{-1}(0)$ intersects $B\cap B'$ at least once.
If it intersects $B\cap B'$ multiple times, we can expand $B$ into
$B^+$ to ensure that $\wtf^{-1}(0)$ intersects the boundary
$\partial_2(B^+)$ exactly once.

For Types (a) and (b), there are two sides of $B'$
where the curve $\wtf^{-1}(0)$ intersects
at least once.  Moreover, there is a unique connected component
$X$ of $\wtf^{-1}(0)\cap (B'\cup B)$ that
connects these two sides (see \refFig{transient}(a,b)).
Recall that we say $X$ has two possible dispositions: either $X$ intersects
the side $B\cap B'$ or it does not.  In either case,
we can expand $B$ into $B^+$ so
that $X$ intersects $\partial_2(B^+)$ exactly twice.
This component $X\cap B^+$
is represented by the augmented edges $(u,v),(v,w)$ in $E^+$.
\epf

We now present the \dt{Extended \PV\ algorithm}.  It has 3 Phases
that parallel the algorithm in Section 2.  Phase $i$ (for $i=1,2,3$)
works off queues $Q_i$ and $Q'_i$, transferring boxes
into $Q_{i+1}$ and $Q'_{i+1}$.  The queue $Q'_i$ holds complementary
boxes while $Q_i$ holds regular boxes.  Initially, $Q_1$ contains $B_0$ and $Q_1'$ contains the four complementary boxes to $B_0$.

\begin{itemize}
\item PHASE 1: SUBDIVISION.  While $Q_1$ is non-empty, remove some $B$
  from $Q_1$, and perform the following: If $C_0(B)$ holds $B$ is discarded.  If $C_1(B)$ holds,
  and also $C_1(B')$ or $C_0(B')$ holds for every complementary box
  $B'$ of $B$, then (a) insert $B$ into $Q_2$, and (b) for each complementary box $B'$
  that satisfies $C_1$ but not $C_0$, insert $B'$ into $Q_2'$.
  Otherwise, split $B$ into four subboxes which are
  inserted into $Q_1$ and \dt{half-split} $B'$
  (this means that we split it into four children and consider the two children
  which intersect $\partial B_0$ and put the two children into $Q_1'$).
\end{itemize}

\begin{itemize}
\item PHASE 2: BALANCING.
  The balancing of boxes in $Q_2$ is done as in Phase 2
  of Section~\ref{basepv}.  Note that boxes are inserted into $Q_3$
  by this process.  Next, we
  perform an analogous while-loop on $Q'_2$:
  While $Q'_2$ is non-empty, remove any $B'$ from $Q_2'$.
  If $B'$ and its partner are different sizes, we then
  \dt{half-split} $B'$ and put the children of $B'$
  into $Q_2'$ provided $C_0$ does not hold.
  Otherwise, we place $B'$ into $Q'_3$.

\item PHASE 3: CONSTRUCTION.
  First, perform Phase 3 of Section~\ref{basepv} which constructs
  a graph $G=(V,E)$ from the boxes in queue $Q_3$.
  Next we augment this graph using queue $Q'_3$:
  for each $B'\in Q'_3$, if $B'$ is a transient type, we insert
  three vertices and two edges to the graph $G$ as described above.
  The resulting straightline graph is denoted $G^+=(V^+,E^+)$.
\end{itemize}

\subsection{Weak Correctness Statement.}\label{WCS}
We are ready to prove the correctness of this Extended \PV\ Algorithm.
We define $B_0'$ to be the union of $B_0$ with all of the
complementary boxes $B'$ that were placed into $Q_3'$.

Before we formulate the correctness statement, we will discuss the implications of using an
infinitesimal perturbation.
In most cases the perturbation does not change the topology,
e.g., when the curve intersects a corner in the interior of $B_0$,
the topology in $B_0$ does not change.
This is the reason that in \citep{plantinga:thesis:06,plantinga-vegter:isotopic:04}, the
curve $\wt{S}=\wtf^{-1}(0)$ is isotopic to $S=f^{-1}(0)$.
In our setting, $\wt{S}\cap B_0$ may no longer be isotopic to $S\cap B_0$.
For instance, if $S$ either makes a non-crossing tangential
intersection with the boundary $\partial B_0$ at the corner of a box $B$ of $T$
or passes through a corner of $B_0$, this
intersection is an isolated component of $S\cap B_0$; we could lose or enlarge
this component in $\wt{S}\cap B_0$, depending on the sign of $f$ nearby.
To correctly analyze this case would require a more delicate consideration of the
bounary, in particular, how complementary boxes of neighbors interact.
Our correctness statement is therefore
about $\wt{S}\cap B_0$, and not about $S\cap B_0$.

We regard the use of $\wtf$ as a reasonable compromise because
(1)
$\wtf$ is an infinitesimal perturbation of $f$;
(2)
$\wtf$ is easy to implement and is an effective perturbation
(comparing favorably to the common alternative of,
say, a randomized perturbation);
(3)
we know exactly when $\wt{S}\cap B_0$ deviates from $S\cap B_0$
(i.e., we encounter a zero at a box corner in the boundary of $B_0$);
(4)
it is a very simple solution to what would otherwise
be serious complications arising from various degeneracies
(e.g., when $S$ contains a horizontal or vertical component);
and finally,
(5)
its use is consistent with the exploitation of isotopy in \PV.

\bthmT{Weak Correctness}{isotopic}
Let the curve $S = f^{-1}(0)$ be non-singular in the box $B_0$,
and $G^+=(V^+,E^+)$ be the augmented straightline graph
constructed by the Extended \PV\ Algorithm.
Then there exists a region $B_0^+$ isotopic to $B_0$,
    $$B_0 \ib B_0^+ \ib B_0',$$
such that
    $$G^+\approx \wtf^{-1}(0) \cap B_0^+.$$
\ethmT
\bpf
Let $T$ be the quadtree at the end of the Extended \PV\ algorithm and
$\wtf$ be the standard perturbation of $f$ relative to $T$.
Using \refThm{normalized}, we have
    \beql{f10}
    \wtf^{-1}(0)\cap B_0  \approx f_1^{-1}(0) \cap B_0,
    \eeql
where $f_1$ is a normalization of $\wtf$ relative to $T$.
Let $B_0^+$ be the union of $B_0$ with all of the caps constructed using \refLem{local}.
Since the isotopy constructed in \refThm{normalized} acts as the identity outside of $B_0$,
it follows from \refeq{f10} that:
    \beql{f1+}
    \wtf^{-1}(0)\cap B_0^+  \approx f_1^{-1}(0) \cap B_0^+.
    \eeql
By a direct consequence of \refThm{normalized},
we have $G^+ \cap B \approx f_1^{-1}(0) \cap B$
for each interior box $B$ of $T$.
For each boundary box $B$ of $T$, \refLem{local} shows that $G^+ \cap B \approx f_1^{-1}(0) \cap B^+$,
where $B^+$ is the union of $B$ and its cap.

Now the composition of all the isotopies involved gives the isotopy
$G^+ \approx f_1^{-1}(0) \cap B_0^+$,
which, when combined with \refeq{f1+}, yields the desired result.
\epf

REMARKS:
\\ 1.
Under the conditions of this theorem,
we call $G^+$ a \dt{weak isotopic approximation}
to $f^{-1}(0)\cap B_0$.
In fact, once there is a collar around $B_0$
where in each complementary box either $C_0$ or $C_1$ holds,
there is a spectrum of similar algorithms that allow for various
correctness statements.
This extension algorithm lies at one end of this spectrum, while the na\"{i}ve extension,
which requires $\wtf^{-1}(0)$ to intersect each boundary segment at most onces lies at the
other end.  The algorithms along this spectrum reflect a trade off between using isotopy on the boundary
and topological correctness within $B_0$.
A slightly different approach, which is based on an algorithm that lies between these two extremes, 
was used in our ISSAC 2008 proceedings version, that solution is in the middle of this spectrum:
there, we required $f^{-1}(0)$ to only intersect $\partial B_0$ transversally.
Then we can explicitly distinguish between
the two dispositions in the transient boxes (Types (a) or (b)).
To do this, we require an iteration at each transient box.
This application is a main reason why these boxes are called transient,
after a finite number of additional subdivisions, the boxes will become non-transient.
Finally, the algorithm in ISSAC 2008
augments $G$ with an edge if and only if this iteration detects an incursion.
We prefer the current solution because it is non-trivial to ensure that
$f^{-1}(0)$ only intersects $\partial B_0$ transversally and because it uses isotopy as much as possible.
\\ 2.
We can make the collar $B'_0\setminus B_0$ around $B_0$ as narrow as desired:
to ensure a perturbation bound of $\vareps>0$, it is sufficient that
each of the complementary boxes where $C_0$ does not hold has width at most $\vareps/4$.
This is easily done by modifying the
balancing phase of the algorithm.
\\ 3.
We have assumed that complementary boxes have
the same width as their partner.
An alternative, possibly more efficient, approach is to allow complementary
boxes to have widths less
than their partners.  Call these ``subcomplementary boxes''.
A boundary box $B$ can have many subcomplementary boxes.  We can do half-splits
of subcomplementary boxes as long as their widths are greater than
$\vareps$.
However, to ensure topological correctness,
it becomes necessary to insist that a box has only a limited number of
certain types of ``subcomplementary boxes''.
For instance, a boundary box $B$ may be restricted to have at most
one subcomplementary box of Types (c) or (d),
and this occurs under strict conditions
(we leave the details to the reader).

\subsection{Extension to Non-simply Connected Regions}

\begin{figure}[htb]
    \begin{center}
	\scalebox{0.3}{
    	  \input{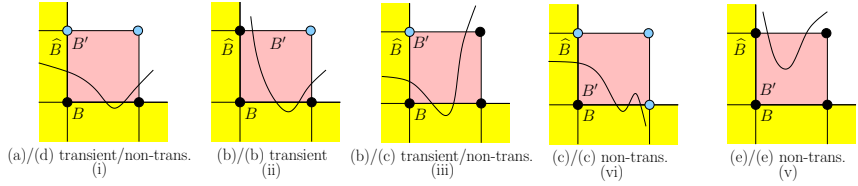}}
    \caption{Classification of complementary boxes
    at concave corners.}
    \label{fig:transient-concave}
    \end{center}
    \end{figure}

It is essential in our applications later
to extend the above refinements to non-simply connected regions.
Recall that we only consider regions $R_0$ which come from subdivisions.
To extend \refThm{isotopic} to these regions,
we note two simple modifications:
\\ (I)
A complementary box $B'$ of
a boundary box $B\ib R_0$ may intersect the interior of
$R_0$ or other complementary boxes.
Thus, Phase 1 must split such boundary boxes $B$ sufficiently.
Such interference checks can be checked during the subdivision phase.
\\ (II)
The region $R_0$ can have concave corners.
A complementary box $B'$ at a concave corner
has two partners $B\ib R_0$.
Relative to each partner $B$, we classify $B'$ into one of 5 types
as in \refFig{transient}(a)-(e).
This is illustrated in \refFig{transient-concave}(i)-(iv).
In \refFig{transient-concave}(i), for instance,
the box $B'$ is Type (a) (hence transient) relative to the indicated box $B$,
but it is Type (d) (hence non-transient)
relative to the other partner $\wh{B}$.
In \refFig{transient-concave}(ii),
the corner complementary box $B'$ is Type (b) (hence
transient) relative to both $B$ and $\wh{B}$.
Similarly, the other two cases
seen in \refFig{transient-concave}(iii)-(iv) have
dual classifications.  We modify our augmentation of
the graph $G=(V,E)$ as follows: for each
complementary box at a concave corner, we consider its
classification relative to each choice of partner $B$:
if the classification is transient, as before,
we add three vertices $u,v,w$ and edges
$(u,v), (v,w)$ to $G$ on the side of $B'\cap B$.

\section{Evaluation Bounds}

For any function $f$, define its {\bf evaluation bound} to be
$$
\EVB(f) \as \inf\{|f(p)| \st f(p)\not=0, \grad f(p)=0\}
$$
Such bounds were used in
\citep{cheng-gao-yap:triangular:07,burr-krahmer-yap:continuousAmort:09}.
From \refPro{sard},
we see that $\set{f(p): \grad f(p)=0}$ is a finite set
and therefore $\EV(f)>0$. However there is no explicit bound
readily available.  The main objective of this section is to
provide such a bound:

\bthml{ev} If $f\in\ZZ[X,Y]$ has degree $d$ and $\|f\|<2^L$ then
$-\lg \EV(f) = O(d^2(L+d\log d))$.  More precisely,
    $$\EV(f)^{-1} \le
        \max\left\{
        [d^6 2^{L + 2d + 11}]^{d^2-1},
        [d^{3d+8} 2^{3L+5d}]^d
        \right\}$$
\ethml

Before giving the proof, we provide some definitions and preparation.
Let $h = \sum_{i+j = 0}^d a_{ij} X^i Y^j \in
\CC[X,Y]$. $\|h\|_k := \sqrt[k]{\sum_{i+j=0}^d |a_{ij}|^k}$ will
denote the $k$-norm of $h$, where we use $k=1,2$. We just
write $\|h\|$ for $\|h\|_\infty := \max|a_{ij}|$, denoting the height of $h$.

Now consider $f$ as in the statement of the theorem
considered as a function on $\mathbb{C}^2$.
As input parameters in our bounds, we use
$d$ and $L$ where $\deg f \le d$ and $\|f\|<2^L$.
Let $f_x, f_y$ denote the derivatives of $f$.
We may write
    $$\V(f_x,f_y) = \bigcup_i U_i \cup \bigcup_j V_j$$
where $U_i$ are the $1$-dimensional irreducible components
and $V_j$ are the $0$-dimensional irreducible components.
On each component $U_i$, one can show that
the function $f$ is constant.
E.g. $f=(xy +1)^2 - 1$, $f_x=2(xy+1)y$ and $f_y=2(xy+1)x$.
Then $U_1=\set{xy+1=0}$ and $V_1=\set{(0,0)}$.
The function $f$ is equal to $-1$ on $U_1$ and $0$ on $V_1$.

Let $g\as \gcd(f_x,f_y)$ and also
    $$\ol{f_x} \as f_x/g, \qquad \ol{f_y} \as f_y/g.$$
Clearly, we have
    $$\V(f_x,f_y)= \V(g)\cup \V(\ol{f_x},\ol{f_y}).$$
Since $\gcd(\ol{f_x},\ol{f_y})=1$,
we conclude that $\V(\ol{f_x},\ol{f_y})$ has no 1-dimensional components.
Conversely, the hyper-surface $\V(g)$ has no $0$-dimensional components,
as a subvariety of $\CC^2$.
This proves:
\bleml{vg}
    $$\V(g)=\bigcup_i U_i,
        \qquad \V(\ol{f_x},\ol{f_y})=\bigcup_j V_j.$$
\eleml

We now view the ring $\ZZ[X,Y]\simeq\ZZ[Y][X]\simeq\ZZ[X][Y]$
in three alternative ways:  A bivariate polynomial $f$ in $X$ and $Y$
can be written as $f=f(X,Y)$, $f=f(Y;X)$ or $f=f(X;Y)$ to indicate these three views respectively.
As a member of $\ZZ[X,Y]$, the coefficients of $f(X,Y)$
are elements of $\ZZ$.  But $f=f(Y;X)$ is
a member of $\ZZ[Y][X]$ whose coefficients
are elements of $\ZZ[Y]$.
The leading coefficient and degree of $f$ are likewise affected by these views:
$\lc(f(Y;X))\in \ZZ[Y]$ but $\lc(f(X,Y))\in\ZZ$,
$d=\deg(f(X,Y))$ is the total degree of $f$
while $\deg(f(Y;X))$ is the largest power of $X$ occurring in $f$.

We use Mahler's basic inequality
(\citep[p.~351]{yap:algebra:bk}) that if $p\in\ZZ[X,Y]$ and $p|q$ then
    \beql{mahler}
    \|p(X,Y)\|_1 \le 2^D \|q(X,Y)\|_1
    \eeql
where $D=\deg(q(X;Y))+\deg(q(Y;X))$.
This implies:
    \beql{g}
    \|g(X,Y)\|_1 \le 4^{d-1} d^3 2^L,
    \qquad
    \|\ol{f_x}(X,Y)\|_1 \le 4^{d-1} d^3 2^L.
    \eeql
since $g|f_x$, $\ol{f_x}|f_x$, and
$\|f_x\|_1  \leq d^2\|f_x\|
\leq d^2 \cdot d 2^L$, $\deg(f_x(X;Y))+\deg(f_x(Y;X))\le 2d-2$.
The bound then follows from \refeq{mahler}.

Let $h(X)$ be the leading coefficient of $g(X;Y)$.
Since $h(X)$ has degree $\le d-1$, there is
an integer $x_0\in\set{0,1\dd d-1}$ such that $h(x_0)\neq 0$.
Intersect $\V(g)$ with the line $X=x_0$.
We claim that this line cuts each non-vertical component $U_i$ in a finite but non-zero
number of points.
In proof, let $g=\prod_i g_i^j$ where $\V(g_i)=U_i$.
Setting $d_i\as \deg g_i(Y;X)$, we see that the vertical components correspond
to $d_i=0$.
Then $\lc(g(X;Y))=\prod_i \lc(g_i(X;Y))$ and
$h(x_0)=\lc(g(x_0;Y))\neq 0$ iff for all $i$, $\lc(g_i(x_0;Y))\neq 0$.
But $g_i(x_0;Y)$ is a polynomial of degree $d_i$ in $\ZZ[Y]$,
and, therefore, has exactly $d_i$ solutions in $\CC$.

Write $f_0(Y)\as f(x_0,Y)$ and $g_0(Y)=g(x_0,Y)$.
From \refeq{g}:
$$
\|g_0\|_1 \le d^d \|g(X,Y)\|_1 \le 4^{d-1} d^{d+3} 2^L.
$$
It is also easy to see that
    $$
    \|f_0\| \le d^{d+1} 2^{L+1}.
    $$
Suppose $\beta\in \V(g_0)\setminus \V(f_0)$.
We want a lower bound on $|f_0(\beta)|$.
For this purpose, we use an evaluation bound from
\citep[Theorem 13(b)]{burr-krahmer-yap:continuousAmort:09}:

\bproT{Evaluation Bound \citep{burr-krahmer-yap:continuousAmort:09}}{eval}
Let $\phi(x), \eta(x)\in\CC[x]$ be complex polynomials
of degrees $m$ and $n$.  Let $\beta_1\dd \beta_n$
be all the zeros of $\eta(x)$.
Suppose there exists relatively prime $F,H\in\ZZ[x]$
such that $F=\phi\ot{\phi}, H=\eta\ot{\eta}$
for some $\ot{\phi},\ot{\eta}\in\CC[x]$.
If the degrees of $\ot{\phi}$ and $\ot{\eta}$ are $\otm$ and $\otn$,
then
    { $$
    \prod_{i=1}^n |\phi(\beta_i)|
    \ge \frac{1}{
        \lc(\ot{\eta})^{m}\cdot
        \left( (m+1)\|\phi\|\right)^{\otn} M(\ot{\eta})^m \cdot
        \left((\otm+1)\|\ot{\phi}\|\right)^{n+\otn}
        M(H)^{\otm}}.
    $$}
\eproT

Here the Mahler measure $M(h)$ of a function $h \in \CC[x]$ with
zeros $\alpha_1,\cdots,\alpha_n$, is defined as
$M(h):=|lc(h)|\prod_{|\alpha_i|\geq 1} |\alpha_i|$.
We shall choose the functions in \refPro{eval}
as follows:
    $$\phi \as f_0,\qquad
      H \as  \frac{g_0}{\gcd(f_0, g_0)}.$$
Moreover, let $\ot{\phi} \as 1$, $\eta(Y) \as Y-\beta$
and $\ot{\eta}\as H/\eta \in \CC[x]$.  Hence
    $$m\le d,\quad
        n=1,\quad \ot{m}=0,\quad \ot{n}\le d-1.$$
Also
    $$
    \lc(\ot{\eta}) = \lc(H) = \lc(g_0) \le \|g_0\|\le \|g_0\|_1.
    $$
Further,
   $$
    M(\ot{\eta}) \le M(H) \le \|H\|_1 \le 2^d \cdot \|g_0\|_1.
    $$
Finally, an application of \refPro{eval} gives
    \beqarrayl{bound1}
    |f_0(\beta)|^{-1} &\le&
        \lc(\ot{\eta})^d \cdot \left((d+1)\|f_0\|\right)^{d-1}
                \cdot M(\ot{\eta})^d \nonumber \\
        &<& \left[ \lc(\ot{\eta}) \cdot d\|f_0\|
                \cdot M(\ot{\eta})
            \right]^d \qquad (\aS\ (d+1)^{d-1}\le d^d \foR\ d\ge 2)\nonumber \\
        &\le& \left[ \|g_0\|_1
            \cdot d d^{d+1} 2^L
            \cdot 2^d\|g_0\|_1
             \right]^d \nonumber \\
        &\le& \left[ 
                d^{3d+8} 2^{3L+5d}
              \right]^d .
    \eeqarrayl
\refeq{bound1} is a lower bound on $|f(p)|$ where $p$ lies in
a non-vertical component $U_i$.   By considering $g(Y;X)$, the same
bound applies for $|f(p)|$ when $p$ lies in a vertical component $U_i$.

Now, we obtain a lower bound for
$f(p)$ with $p\in \V(\ol{f_x},\ol{f_y})$.
Consider the system $\Sigma\ib \ZZ[X,Y,Z]$ where
    $$\Sigma =\set{Z-f(X,Y), \ol{f_x}(X,Y), \ol{f_y}(X,Y)}$$
The zeros $(\xi_1,\xi_2,\xi_3)=(x,y,f(x,y))\in\CC^3$ of $\Sigma$ satisfy
$\xi_3=f(\xi_1,\xi_2)$.
Since $\Sigma$ is a zero dimensional system, we may apply the multivariate
zero bound in \citep[p.~350]{yap:algebra:bk}.  This bound says that
    $$
    |\xi_3|^{-1} < (2^{3/2} N K)^{D} 2^{8(d-1)}
    $$
where $N={1+2(d-1)\choose 3}$, $D=d^2-1$ and
$$K=\max\set{\sqrt{3},\|\ol{f_x}\|_2, \|\ol{f_y}\|_2,\|Z-f(X,Y)\|_2}.$$
We have $\|Z-f(X,Y)\|_2\le (d+1)2^{L+1}$.
From \refeq{g}, we see that $K\le 4^dd^32^L$.
Using the bound $N<2d^3$, we obtain
    \beql{bound3}
    |\xi_3|^{-1} < [2^2 \cdot 2d^3 \cdot 4^d d^3 2^L]^{d^2-1} \cdot 2^{8(d^2-1)}
                 < [d^6 2^{L + 2d + 11}]^{d^2-1}.
    \eeql
Now \refThm{ev} easily follows from \refeq{bound1} and
\refeq{bound3}.

\section{Isolating Singular Points}

In the remainder of this paper, we assume that $f\in\ZZ[X,Y]$
and allow the curve $S=f^{-1}(0)$ to have singular points,
except on the boundary $\partial B_0$.
We would like to
use the Extended \PV\ algorithm to compute an isotopic
approximation to $\V(f)$ when $f$ has only isolated
singularities. Since the \PV\ algorithm does not terminate near
singular points, it is necessary to isolate the singular points
from the rest of $B_0$.

We use the auxiliary function $F =
f^2+f_X^2+f_Y^2$. Finding the singular points of $f^{-1}(0)$ amounts
to locating and isolating the zeros of this non-negative function. We use a simple mountain pass theorem
\citep{jabri:mountain-pass:bk} adapted to $B_0$ to ensure our algorithm isolates the zeros.

\bthml{mountain} Suppose that $F\geq0 $ on $B_0$, and that $F > 0$
on $\partial B_0$. Then for any two distinct roots $p, q$ of $F$
in $B_0$, there exists a continuous path $\gamma : [0,1] \to B_0$
connecting $p$ and $q$ which satisfies the following:

\begin{itemize}
\item $\gamma$ minimizes $M_\gamma := \max_{x\in [0,1]}
F(\gamma(x))$ among all paths connecting $p$ and $q$ in $B_0$.
\item $\gamma$ contains a point $y$ such
that either $\nabla F(\gamma(y)) = 0$ or $\gamma(y) \in \partial
B$.
\end{itemize}

\ethml

This can be proved using path deformation and the compactness of
$B_0$, or it can be seen as a simple application of the
topological mountain pass theorem presented in
\citep{jabri:mountain-pass:bk}. 
Because of this theorem, distinct zeros of $F$ within $B_0$ are
separated by barriers of height $\epsilon = \min(\EVB(F), \min
F(\partial B_0))$. This leads us to the following multistep process to
localize these zeros.  The goal is to find a small rectangle with
diameter less than some $\delta$ around each zero.

STEP 0: DETERMINING $\epsilon$.
Initialize $\epsilon$ to any lower bound on $\EVB(F)$.
Also, initialize $Q_0$ to be $\set{B_0}$ and $Q_1$ to be empty.
$Q_0$ will contain boxes $S$ that intersect $\partial B_0$ and
have $0\in\intbox F(S)$; these boxes will then be subdivided in order to
compute a lower bound on $F(\partial B_0)$.
While $Q_0$ is non-empty, remove a square $S$ from it and evaluate
$\intbox F(S)$. If $\intbox F(S) >0$ we push $S$ into queue
$Q_1$ and also update $\epsilon$ to $\min\set{\epsilon,\min\intbox
F(S)}$. If $0\in\intbox F(S)$, subdivide $S$ and push the children
of $S$ which intersect $\partial B_0$ into $Q_0$, and the others
into $Q_1$. When $Q_0$ is empty, we stop and fix the current
value of $\epsilon$ for the remainder of the algorithm.

STEP 1: INITIAL SUBDIVISION.
Initialize queue $Q_2$ to be empty.
While there is an $S$ in $Q_1$, remove it and evaluate $\intbox F(S)$.
If $\intbox F(S) > \epsilon/2$, then discard $S$.  Else if
$\intbox F(S) < \epsilon$, push $S$ into $Q_2$.
Else subdivide $S$ and push its children into $Q_1$.

Once $Q_1$ is empty, group the elements of $Q_2$ into connected
regions $A_i$ ($i\in I$).
Each $A_i$ contains at most one root, since otherwise,
there would be a path connecting the roots within $A_i$. The value of $F$ along
this path would be less than $\epsilon$,
contradicting the mountain pass theorem.  Let $C$ be the region
$B_0\setminus \cup_i A_i$.  Note by Step 0 that $F$ is greater than $\epsilon/2$
on $C$ and that $\partial B_0 \ib C$, i.e., each $A_i$ doesn't intersect $\partial B_0$.

STEP 2: REFINEMENT.
For each $A_i$ ($i\in I$), initialize queue
$Q_{2,i}$ with all squares $S\in A_i$.
So long as neither terminating condition 1 nor 2 (below) hold, we perform the
following: For each $S$ in $Q_{2,i}$, if $0\in\intbox F(S)$, subdivide
$S$ and push its children into $Q_{2,i}$.  If $0\not\in \intbox F(S)$, discard
$S$.  We terminate when either of the following two conditions are met:
\begin{enumerate}
\item
$Q_{2,i}$ is empty, in which case
there isn't a zero in $A_i$.
\item
$A'_i$, the contents of $Q_{2,i}$ satisfy all of the following:
\begin{enumerate}
 \item
 $\intbox F(S) < \epsilon/2$ for some $S\in A'_i$
\item
 $R_i$, the smallest rectangle containing $A'_i$, lies within
the region covered by the original $A_i$.
\item
 The diameter of $R_i$ is less than $\delta$.
\end{enumerate}
\end{enumerate}

It is clear from the definition of $F$ that this step will halt.
We claim that each $R_i$ contains exactly one root. In
Step 1, we showed that $A_i$ contains at most one root.
To see that $R_i$ contains a
root, take a point of $A'_i$ where
$\epsilon/2$, 
then follow the path of steepest descent to reach a
zero of $F$. Because $F$ is less than $\epsilon/2$
on this curve, the
curve cannot pass through the region $C$ to reach any other $R_j$
or to leave $B_0$. Therefore there must be a zero within $A_i$.
It is in $R_i$ because our conditions
ensure that $F$ is positive on $A_i\setminus R_i$.

\section{Determining the Degree of Singular Points}
\label{section:degree}
The following standard result from \citep{krantz-parks:primer:bk,lojas:bk}
shows that the global structure of zero sets:

\bproT{Zero Structure}{structure}
Let $f$ be real analytic.
Then $\V(f)$ can be decomposed into a finite union of pieces homeomorphic
to $(0,1)$, pieces homeomorphic to $S^1$, and singular points.
\eproT

\noindent In our current situation, the pieces which are homeomorphic to $(0,1)$
are smooth open subsets of the irreducible components of $\V(f)$.

Viewing $\V(f)$ as a multigraph $H$,
the branching \dt{degree} of a singular point is its
degree as a vertex of $H$.  We now determine such degrees.
Let $\delta_3$ be a separation bound between singular points,
so if $p$ and $q$ are two distinct singular
points of $\V(f)$, then the distance between $p$ and $q$
is at least $\delta_3$.
Let $\delta_4$ be
a separation bound so that if $r$ is a point on $\V(f)$
such that $\grad f(r)$ is parallel to the line between $r$
and a singular point $p$, then the distance between
$p$ and $r$ is at least $\delta_4$.
If $s$ is on $\V(f)$ so that the distance between $s$ and a singular point
$p$ is smaller than either $\delta_3$ or $\delta_4$, then
by following the paths $\V(f)$ away from $s$, one of
the paths strictly monotonically approaches $p$ until it
reaches $p$ and the other path locally strictly monotonically recedes
from $p$.
\citep{yap:subdiv1:06} provided an explicit bound on $\delta_3$
as a function of degree and height of $f(X,Y)$:
    $$\delta_3 \ge \min\left\{(16^{d+2} 256^L 81^{2d} d^5)^{-d},(2^{8L+21}3^{8d})^{-2}\right\}.$$

\subsection{Lower Bound on $\delta_4$.}
To derive an explicit bound on $\delta_4$, we
consider the following $6$ polynomials in $\ZZ[z,p_x,p_y,q_x,q_y]$:
    \beql{if}
      I_f\as \set{
        f(p), f(q), f_x(q), f_y(q), (p-q)_x f_y(p)-(p-q)_y f_x(p), z^2-\|p-q\|^2}
    \eeql
where $p=(p_x,p_y)$ and $q=(q_x,q_y)$ are points on $\V(f)$.
$q$ is a singular point of $\V(f)$.
Moreover, define ``$(p-q)_x$'' to mean $p_x-q_x$, so that the equation
$(p-q)_x f_y(p)-(p-q)_y f_x(p)=0$
implies that $(p-q)$ is parallel to $\grad f(p)$.
The equation $z^2-\|p-q\|=0$ implies that $z$
is the distance between $p$ and $q$.

Consider the projection $\Pi_z[\V(I_f)]$ of the zeros of
$I_f$ onto the $z$-coordinate.  Then
$\delta_4$ is $\inf\set{|z|: z\in \Pi_z[\V(I_f)], z\neq 0}$.
We obtain a lower bound on $\delta_4$ using
the following general theorem from
\citep{brownawell-yap:bound:09}:

\bprol{brownawell-yap}
  Let $I \as (P_1,\dd, P_m) \ib A \as \ZZ[X_1,\dots,X_n]$. Let $\fP$ be
  an isolated prime component of $I$ whose projection
  onto the first coordinate, $\Pi_1(\V(\fP))$, is a finite
  set.  If $\olzeta=(\zeta_1,\dd, \zeta_n)\in \V(\fP)$ and
  $\zeta_1\neq 0$, then
\[
\label{E:main}
        |\zeta_1| \ge
((n+1)^2e^{n+2})^{-n(n+1)D^{n-k}} (k^{n-k-1}mH)^{-(n-k)D^{n-k-1}},
\]
where
\begin{itemize}
\item $\dim \fP = k$,
\item $H\ge \Height(P_i)$, and
\item  $D\ge \deg(P_i)$, $i=1,\dd, m$.
\end{itemize}
\eprol

If $\Pi_z[\V(I_f)]$ is a finite set,
then we use the bounds $n=5$, $m=6$, $k\leq 4$, $D=\max\{2,d\}$, and $H\leq d2^{L+1}$ in applying this theorem to get the following bound:
$$\delta_4 \ge (6^2e^7)^{-30D^5} (4^4 \cdot 6\cdot d2^{L+1})^{-5D^4}$$
Combining the two cases for the value of $D$ gives
$$\delta_4 \ge \min\left\{ (6^2e^7)^{-30d^5} (4^4 \cdot 6\cdot d2^{L+1})^{-5d^4},(6^2e^7)^{-30\cdot2^5} (4^4 \cdot 6\cdot d2^{L+1})^{-5\cdot2^4}\right\}$$
It remains to show that $\Pi_z[\V(I_f)]$ is a finite set.
We prove this in the following lemma:

\bleml{sphere}
$\Pi_z[\V(I_f)]$ is a finite set.
\eleml
\bpf
Without loss of generality,
we apply a translation so that we can assume that $q$ is at the origin.
To show that this image is a finite set, we show that
$\V(f(p),p_xf_y(p)-p_yf_x(p))$ is contained in finitely many circles centered
at the origin.
Then, the possible values of $z$ are the radii of these circles,
of which there are finitely many.

By \refPro{structure},
we know that each component of $\V(f)$ is either a single point
or a smooth one dimensional manifold without boundary.
Since there are finitely many components which are a single point,
these components are contained within finitely many circles centered at the origin.

Take any one dimensional component $M$ and let $r,s$ be two points in $M$ and $\gamma:[0,1]\to M$ a smooth path from $\gamma(0)=r$ to $\gamma(1)=s$ in $M$.  Also, define $\rho : \CC^2 \to \CC$ by $\rho(x,y) = x^2+y^2$.
For any $p\in M$, we note that since $M$ is smooth,
the tangent line of $M$ (or equivalently $\V(f)$) at $p$ is perpendicular to $\nabla(f)=(f_x(p),f_y(p))$ and
this gradient is non-zero because $p$ is a smooth point of $\V(f)$.
In addition, since $p_xf_y(p)-p_yf_x(p)$,
we know that $(p_x,p_y)\parallel (f_x(p),f(y))$.
Also note that $\grad \rho=(2x,2y)=2\cdot (x,y)$.

Now, we consider the square of the distance between $\gamma$ and the origin, $\rho(\gamma(t))$.
Taking the derivative of this function gives
	$$\frac{d}{dt}\rho(\gamma(t))
		=(\nabla\rho)(\gamma(t))\cdot\gamma'(t)
		=2\gamma(t)\cdot\gamma'(t) = 0.$$
Since this derivative is always zero, it implies the distance from the
origin to points $\gamma(t)$ on $M$ is constant.
Therefore, $M$ is contained in a unique circle centered at the origin.
\epf

\begin{figure}[htb]
    \begin{center}
	\scalebox{0.5}{
    	  \input{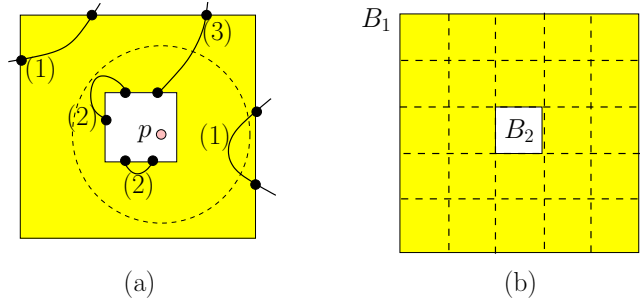}}
    \caption{Annular region $B_1\setminus B_2$ with singularity $p$
    and the three types (1), (2), (3) of components.}
    \label{fig:radial}
    \end{center}
    \end{figure}

To find the degree of a singular point,
assume that we have two boxes $B_1\varsupsetneq B_2$ where
the diameter of $B_1$ is less than both $\delta_3$ and
$\delta_4$, $B_2$ contains a singular point of $f$ and there is
some radius $r>0$ such that a circle of
radius $r$ centered at any point $p$ inside $B_2$ must lie entirely
within the annulus $B_1\setminus B_2$, see \refFig{radial}(a).
Note this condition is satisfied if $B_1$ is at least 5 times larger than $B_2$
and $B_2$ is centered in $B_1$, which is the typical situation we consider below.
See \refFig{radial}(b).
Furthermore, to apply our extended \PV\ algorithm of Section 4,
we ensure that $B_1\setminus B_2$ comes from a subdivision.

Now, there are 3 types of components (other than isolated points) in
$\V(f)\cap (B_1\setminus \interior(B_2))$:
(1) images of $[0,1]$ both of whose endpoints
are on $\partial B_1$, (2) images of $[0,1]$ both of whose
endpoints are on $\partial B_2$, and (3) images of $[0,1]$
with one endpoint on each of $\partial B_1$ and
$\partial B_2$.  These three types are illustrated in \refFig{radial}(a).

Let $s$ be a point on {\em any} of
these components, then traveling along $\V(f)$ in one direction
must lead to the singular point and the other
direction must leave the neighborhood (be further than
$\min\{\delta_3,\delta_4\}$) of the singular point.
For, if not, then there must be a point $r$ on $\V(f) \cap B_1$ such that
$\nabla f(r)$ is in the same direction as the line between $s$ and the singular point,
which is impossible since the width of $B_1$ is smaller than $\delta_4$.
Now, any piece defined by \refPro{structure} which reaches the
singular point exits the neighborhood of the singular point and the
only way to leave the neighborhood is by way
of a type (3) component.
In addition, each piece includes either one or two components of type (3), and it only includes two components if both endpoints reach the singular point. This follows from the choice of $\delta_4$.
This shows:

\blem
The degree of the singular point in $B_2$ is the
number of components of type 3.
\elem

\section{Overall Algorithm}

We now put all the above elements together to find
a weak isotopic approximation
to the algebraic curve $S=f^{-1}(0)$
within a region $R_0$, coming from a subdivision,
where $f(X,Y)\in\ZZ[X,Y]$ has
only isolated singularities.
We first find the singularities of the curve $S$
in $R_0$.  Using the technique of Section 5, we can
isolate the singularities $p_i$ ($i=1,2,\ldots$) into
disjoint boxes $B_i$.  We assume the width of
the $B_i$'s is at most $\min\set{\delta_3,\delta_4}/6$.
Let $B'_i$ be the box of width $5$ times
the width of $B_i$, and concentric with $B_i$; we further
assume $B'_i\ib R_0$. Note that these combinations are chosen to ensure that we have the typical
situation in Section~\ref{section:degree}.
Now we proceed to run the extended \PV\ algorithm on the nice
region $R^*\as R_0\setminus \bigcup_i B_i$, yielding a polygonal
approximation $G$.
We directly incorporate the technique of Section~\ref{section:degree} into the
following argument.  If $p_i$ is the
singular point in $B_i$, then the degree of $p_i$ is equal to
the number of type (3) components in $G\cap (B'_i\setminus B_i)$.
We connect these components directly to $p_i$,
and discard any type (2) components.
This produces the desired isotopic approximation.

Remarks:
We have not discussed $\vareps$-approximation because
this is relatively easy to achieve in the \PV\ approach.
We only have to make sure that each subdivision box that contains
a portion of the polygonal approximation $G$ has width
at most $\vareps/4$ since the result of the \PV\ algorithm only deforms the original
curve at most one cell away.

\section{Conclusion}

This paper presents the first
complete numerical subdivision algorithm for meshing an implicit
algebraic curve that has only isolated singularities.
This solves an open problem in the exact numerical approaches
to meshing in 2-D \citep[p.~187]{boissonnat+4:meshing:06}.
We pose three challenges:
\\(a)
A worst case complexity bound for our procedure is
possible.  But this may not be the best way
to measure adaptive algorithms.
We would like to provide adaptive bounds,
similar to the integral analysis in
\citep{burr-krahmer-yap:continuousAmort:09} for 1-D problems.
\\(b)
In 3-D, a square-free integer polynomial $f(X,Y,Z)$
could have a 1-dimensional singularities.
We pose the problem of designing a purely
numerical subdivision algorithm to handle 1-dimensional singularities.
\\(c)
The practical implementation of an adaptive algorithm
handling singularities, even based on our outline,
must handle many important details.
Computational experience is invaluable for
future research into singularity computation.

\bibliographystyle{plainnat}
\bibliography{st,yap,exact,alge,mesh}

\begin{thebibliography}{29}
\providecommand{\natexlab}[1]{#1}
\providecommand{\url}[1]{\texttt{#1}}
\expandafter\ifx\csname urlstyle\endcsname\relax
  \providecommand{\doi}[1]{doi: #1}\else
  \providecommand{\doi}{doi: \begingroup \urlstyle{rm}\Url}\fi

\bibitem[Basu et~al.(2003)Basu, Pollack, and Roy]{basu-pollack-roy:bk}
Saugata Basu, Richard Pollack, and Marie-Fran\c{c}oise Roy.
\newblock \emph{Algorithms in Real Algebraic Geometry}.
\newblock Algorithms and Computation in Mathematics. Springer, 2003.

\bibitem[Boissonnat and Oudot(2006)]{boissonnat-oudot:lipschitz:06}
J.-D. Boissonnat and S.~Oudot.
\newblock Provably good sampling and meshing of {L}ipschitz surfaces.
\newblock In \emph{Proc. 22nd ACM Symp. on Comp. Geometry}, pages 337--346,
  2006.
\newblock Sedona, Arizona.

\bibitem[Boissonnat et~al.(2006)Boissonnat, Cohen-Steiner, Mourrain, Rote, and
  Vegter]{boissonnat+4:meshing:06}
J.-D. Boissonnat, D.~Cohen-Steiner, B.~Mourrain, G.~Rote, and G.~Vegter.
\newblock Meshing of surfaces.
\newblock In J.-D. Boissonnat and M.~Teillaud, editors, \emph{Effective
  Computational Geometry for Curves and Surfaces}. Springer, 2006.
\newblock Chapter 5.

\bibitem[Boissonnat et~al.(2004)Boissonnat, Cohen-Steiner, and
  Vegter]{boissonnat+2:meshing-topology:04}
Jean-Daniel Boissonnat, David Cohen-Steiner, and Gert Vegter.
\newblock Isotopic implicit surfaces meshing.
\newblock In \emph{ACM Symp. Theory of Comput.}, pages 301--309, 2004.

\bibitem[Brownawell and Yap(2009)]{brownawell-yap:bound:09}
W.~D. Brownawell and Chee~K. Yap.
\newblock Lower bounds for zero-dimensional projections.
\newblock In \emph{Proc.~34th Int'l Symp. Symbolic and Algebraic Comp.
  (ISSAC'09)}, pages 79--86, 2009.
\newblock KIAS, Seoul, Korea, Jul 28-31, 2009. Submitted, JSC.

\bibitem[Burr et~al.(2009{\natexlab{a}})Burr, Krahmer, and
  Yap]{burr-krahmer-yap:continuousAmort:09}
Michael Burr, Felix Krahmer, and Chee Yap.
\newblock Continuous amortization: A non-probabilistic adaptive analysis
  technique.
\newblock \emph{Electronic Colloquium on Computational Complexity (ECCC)},
  TR09\penalty0 (136), December 2009{\natexlab{a}}.
\newblock URL \url{http://eccc.hpi-web.de/report/2009/136/}.

\bibitem[Burr et~al.(2009{\natexlab{b}})Burr, Sharma, and
  Yap]{burr-sharma-yap:eval:09}
Michael Burr, Vikram Sharma, and Chee Yap.
\newblock Evaluation-based root isolation, February 2009{\natexlab{b}}.
\newblock In preparation.

\bibitem[Cheng et~al.(2007)Cheng, Gao, and Yap]{cheng-gao-yap:triangular:07}
Jin-San Cheng, Xiao-Shan Gao, and Chee~K. Yap.
\newblock Complete numerical isolation of real zeros in general triangular
  systems.
\newblock In \emph{Proc.~Int'l Symp. Symbolic and Algebraic Comp. (ISSAC'07)},
  pages 92--99, 2007.
\newblock Waterloo, Canada, Jul 29-Aug 1, 2007. DOI:
  http://doi.acm.org/10.1145/1277548.1277562. In press, Journal of Symbolic
  Computation.

\bibitem[Cheng et~al.(2004)Cheng, Dey, Ramos, and Ray]{cheng+3:sampling:04}
S.-W. Cheng, T.K. Dey, E.A. Ramos, and T.~Ray.
\newblock Sampling and meshing a surface with guaranteed topology and geometry.
\newblock In \emph{Proc. 20th ACM Symp. on Comp. Geometry}, pages 280--289,
  2004.

\bibitem[Cox et~al.(1992)Cox, Little, and O'Shea]{clo1:bk}
D.~Cox, J.~Little, and D.~O'Shea.
\newblock \emph{Ideals, Varieties and Algorithms: An Introduction to
  Computational Algebraic Geometry and Commutative Algebra}.
\newblock Springer-Verlag, New York, 1992.

\bibitem[Du and Yap(2006)]{du-yap:hyper:06}
Zilin Du and Chee Yap.
\newblock Uniform complexity of approximating hypergeometric functions with
  absolute error.
\newblock In {Sung-il} Pae and {Hyungju} Park, editors, \emph{Proc.~7th Asian
  Symp.~ on Computer Math.~(ASCM 2005)}, pages 246--249, 2006.

\bibitem[Harris(1992)]{harris:alge-geom:bk}
Joe Harris.
\newblock \emph{Algebraic Geometry}.
\newblock Springer-Verlag, New York, 1992.

\bibitem[Hartshorne(1977)]{hartshorne:bk}
Robin Hartshorne.
\newblock \emph{Algebraic Geometry}.
\newblock Springer-Verlag, New York, 1977.

\bibitem[Hong(1996)]{hong:plane-curves:96}
H.~Hong.
\newblock An efficient method for analyzing the topology of plane real
  algebraic curves.
\newblock \emph{Mathematics and Computers in Simulation}, 42:\penalty0
  571--582, 1996.

\bibitem[Jabri(2003)]{jabri:mountain-pass:bk}
Youssef Jabri.
\newblock \emph{The Mountain Pass Theorem: Variants, Generalizations and Some
  Applications}.
\newblock Encyclopedia of Mathematics and its Applications. Cambridge
  University Press, 2003.

\bibitem[Krantz and Parks(1992)]{krantz-parks:primer:bk}
S.~G. Krantz and H.~R. Parks.
\newblock \emph{A primer of real analytic functions}.
\newblock Birkh\"auser Verlag, Basel, 1992.
\newblock ISBN 3-7643-2768-5.

\bibitem[Lin and Yap(2009)]{lin-yap:cxy:09}
Long Lin and Chee Yap.
\newblock Adaptive isotopic approximation of nonsingular curves: the
  parametrizability and non-local isotopy approach.
\newblock In \emph{Proc.~25th ACM Symp. on Comp. Geometry}, pages 351--360,
  June 2009.
\newblock Aarhus, Denmark, Jun 8-10, 2009. Accepted for Special Issue of SoCG
  2009 in DCG.

\bibitem[{\L}ojasiewicz(1991)]{lojas:bk}
S.~{\L}ojasiewicz.
\newblock \emph{Introduction to complex analytic geometry}.
\newblock Birkh\"auser Verlag, Basel, 1991.
\newblock ISBN 3-7643-1935-6.
\newblock Translated from the Polish by Maciej Klimek.

\bibitem[Lorensen and Cline(1987)]{marching-cube}
W.~E. Lorensen and H.~E. Cline.
\newblock Marching cubes: {A} high resolution 3{D} surface construction
  algorithm.
\newblock In Maureen~C. Stone, editor, \emph{Computer Graphics (SIGGRAPH '87
  Proceedings)}, volume~21, pages 163--169, July 1987.

\bibitem[Mourrain and T{\'e}court(2005)]{mourrain-tecourt:surface:05}
Bernard Mourrain and J.-P. T{\'e}court.
\newblock Isotopic meshing of a real algebraic surface.
\newblock Technical Report RR-5508, INRIA, Sophia-Antipolis, France, February
  2005.
\newblock Also, electronic proceedings, MEGA 2005.

\bibitem[Plantinga(2006)]{plantinga:thesis:06}
Simon Plantinga.
\newblock \emph{Certified Algorithms for Implicit Surfaces}.
\newblock {Ph.D.} thesis, Groningen University, Institute for Mathematics and
  Computing Science, Groningen , Netherlands, December 2006.

\bibitem[Plantinga and Vegter(2004)]{plantinga-vegter:isotopic:04}
Simon Plantinga and Gert Vegter.
\newblock Isotopic approximation of implicit curves and surfaces.
\newblock In \emph{Proc. Eurographics Symposium on Geometry Processing}, pages
  245--254, New York, 2004. ACM Press.

\bibitem[Schoemer and Wolpert(2006)]{schoemer-wolpert:quad:06}
Elmar Schoemer and Nicola Wolpert.
\newblock An exact and efficient approach for computing a cell in an
  arrangement of quadrics.
\newblock \emph{Comput. Geometry: Theory and Appl.}, 33:\penalty0 65--97, 2006.

\bibitem[Seidel and Wolpert(2005)]{seidel-wolpert:topology:05}
Raimund Seidel and Nicola Wolpert.
\newblock On the exact computation of the topology of real algebraic curves.
\newblock In \emph{Proc. 21st ACM Symp. on Comp. Geometry}, pages 107--116,
  2005.
\newblock Pisa, Italy.

\bibitem[Snyder(1992{\natexlab{a}})]{snyder:generative:bk}
J.~M. Snyder.
\newblock \emph{Generative Modeling for Computer Graphics and {CAD}: Symbolic
  Shape Design using Interval Analysis}.
\newblock Academic Press, 1992{\natexlab{a}}.

\bibitem[Snyder(1992{\natexlab{b}})]{snyder:interval:92}
J.~M. Snyder.
\newblock Interval analysis for computer graphics.
\newblock \emph{SIGGRAPH Comput.Graphics}, 26\penalty0 (2):\penalty0 121--130,
  1992{\natexlab{b}}.

\bibitem[Stander and Hart(1997)]{stander-hart:topology:97}
Barton~T. Stander and John~C. Hart.
\newblock Guaranteeing the topology of an implicit surface polygonalization for
  interactive meshing.
\newblock In \emph{Proc. 24th Computer Graphics and Interactive Techniques},
  pages 279--286, 1997.

\bibitem[Yap(2000)]{yap:algebra:bk}
Chee~K. Yap.
\newblock \emph{Fundamental Problems of Algorithmic Algebra}.
\newblock Oxford University Press, 2000.

\bibitem[Yap(2006)]{yap:subdiv1:06}
Chee~K. Yap.
\newblock Complete subdivision algorithms, {I}: Intersection of {B}ezier
  curves.
\newblock In \emph{22nd ACM Symp. on Comp. Geometry}, pages 217--226, July
  2006.

\end{thebibliography}

\end{document}